\documentclass[preprint,floatfix] {revtex4} 
\newcommand{\rvec}{\mathrm {\mathbf {r}}} 
\newcommand{\rpvec}{\mathrm{\mathbf{r'}}}
\usepackage{graphicx}
\usepackage{subfigure}
\begin{document}

\title{A general method for central potentials in quantum mechanics}
\author{Amlan K.\ Roy}
\email{akroy@iiserkol.ac.in, akroy@chem.ucla.edu} 
\affiliation{Division of Chemical Sciences, Indian Institute of Science Education and \\
Research (IISER), Block FC, Sector III, Salt Lake, Kolkata-700106, India.}

\begin{abstract}
This article focuses on a recently developed generalized pseudospectral method for accurate, efficient 
treatment of certain central potentials of interest in various branches in quantum mechanics, usually 
having singularity. Essentially this allows \emph{optimal, nonuniform} spatial discretization of the 
pertinent single-particle Schr\"{o}dinger equation satisfying Dirichlet boundary condition leading to 
standard diagonalization of symmetric matrices. Its validity and feasibility have been demonstrated 
for a wide range of important potentials such as Hulth\'en, Yukawa, generalized spiked harmonic oscillators, 
Hellmann, Coulomb potentials without/with various perturbations (for instance, linear and quadratic) etc. 
Although initially designed for \emph{singular} potentials, this has also been remarkably successful 
for various other cases such as power-law, logarithmic, harmonic potentials containing higher order 
perturbations, 3D rational potentials as well as confinement studies. Furthermore, a large number of 
low-, moderately high-, high-lying multiply excited Rydberg states such as singly, doubly excited He as well as 
triply excited \emph{hollow} $2l2l'2l'' (n \ge 2)$ and $3l3l'3l''$ \emph{doubly-hollow} resonances in 
many-electron atoms have been treated by this approach within a Hohenberg-Kohn-Sham density functional 
theory with great success. This offers very high-quality results for \emph{both ground and higher lying} 
states for arbitrary values of potential parameters (covering both \emph{weak and strong} coupling) with 
equal ease and efficacy. In all cases, excellent agreement with literature results are observed; in many 
cases this surpasses the accuracy of all other existing results while in other occasions our results are 
comparable to the best ones available in literature. This provides a simple general efficient route 
towards the understanding of a number of singular and other potentials of relevance in quantum mechanics 
including multiply excited Rydberg series of many-electron systems. This helps predicting many new states 
for the first time.
\end{abstract}
\maketitle

\section{background and motivation}
Study of singular potentials in quantum mechanics is almost as old as quantum mechanics itself. Many 
important areas in physics and chemistry, such as atomic, molecular, solid-state, nuclear and particle 
physics, field theory, astrophysics, etc., frequently demand quantum mechanical solutions where the 
governing Hamiltonian contains certain central potential (typically having a \emph{singularity}). Often 
this also includes an extra external perturbation term characterizing the physical system under 
investigation. \emph{Exact} analytical solution of the respective Schr\"{o}dinger equation could be 
obtained only for a handful of idealized, model situations, such as the harmonic oscillator or Coulomb 
potential, which are unfortunately quite inadequate for majority of our realistic problems. Thus,
leaving aside a very few privileged cases, for almost all practical purposes, recourse must be taken to 
approximation methods. Consequently, an impressive amount of approximate analytical as well as numerical 
methods have been proposed over the decades, employing a variety of attractive elegant techniques for 
their studies. In general, singular potentials pose more spectacular difficulties and challenges than the regular 
ones. Therefore, development of a \emph{general} method which can offer accurate reliable results on such 
potentials has constituted one of the most fruitful and active areas of research for long time, and 
still this continues to grow in time.

Because of the difficulties concerning physical interpretation of attractive singular potentials, 
scientific community concluded that no significance could be attached to any singular potential as regards 
singularity at the center of force. Also, the mathematical difficulties in tackling these potentials 
made these more formidable. Historically, probably the first  important observation was made for 
well-known Coulomb potentials in a relativistic case. Plesset \cite{plesset32}, while investigating the 
Dirac equation for an electron in a Coulomb field of a fictitious nucleus of charge $\alpha Z \! > \! 137$, 
surprisingly found that the essential distinction between attractive and repulsive potential was lost; 
more precisely all potentials tend to display characteristic features of an attractive potential near 
the origin. They both produce large and small components of wave function behaving in that region 
like a power of $r$ times a factor, $$ \exp \left[ \pm i \int^r V(r)\ dr \right], $$ which is in sharp 
contrast with the non-relativistic case, where $V(r)$ rather than $\sqrt{V(r)}$ appears in the exponential. 
The Klein-Gordon equation also shows same behavior. Non-uniqueness of the solution was also noted by Case 
\cite{case50}, who resolved this dilemma by specifying one bound-state energy, and determining the rest of the 
bound-state spectrum by imposing orthogonality on wave functions. 

In another stimulating paper \cite{predazzi62} in this direction, the authors argued that physical 
interactions in real-world problems were more likely to be highly singular rather than regular 
(non-singular), and naturally their studies were more relevant than their counterparts. They also showed 
that singular potentials display Regge behavior in more simple terms than the regular potentials. Finding 
effective potentials for field-theoretical interactions by means of Bethe-Salpeter or quasi-potential 
equations has given further impetus to the subject. At the same time, in a pioneering work \cite{bastai63}, 
a correlation between renormalizability attributes of a field theory and nature of the effective potential 
was established. The effective potential for super-normalizable, renormalizable and non-renormalizable 
theories were found to be regular, transitional and singular respectively. Further work on peratization 
approximation in the context of field theoretical studies of weak interactions \cite{feinberg63,feinberg64} 
has generated much interest in the study of singular potentials. In the elementary particle scattering, 
short-range interactions between such particles are described by repulsive singular potentials. In molecular 
physics as well, long-range and short-range part of the inter-atomic and inter-molecular forces are 
represented by various singular potentials. The long-range part includes purely electrostatic forces between polar 
molecules or induction forces between a polar and non-polar molecule or dispersion forces between two 
non-polar molecules; corresponding potential is attractive singular when extended to the origin. 
Whereas the short-range force develops due to an overlap of their electron clouds as atoms or 
molecules approach each other at shorter distances. This force is typically represented phenomenologically 
by a repulsive singular (or sometimes non-singular) potential, such as a Lennard-Jones potential. 

The preceding examples illustrate some of the broad application areas where singular potentials serve as 
mathematical models for certain concepts (underlying interaction forces). Time is ripe now to mention a 
few words about some of the specific potentials. The literature is vast and here we restrict ourselves to 
only a few selective ones. One of the deceptively simple one-dimensional potentials $V(x) \! = \! -e^2/|x|$ has 
been studied by many workers, \cite{loudon59,elliott60,cole69,andrews76,davtyan87,nunez87,fisher95,
kurasov96,fisher97,gordeyev97,li01,palma06} mainly because of following reasons: (a) exact solvability 
(b) unfortunate formal resemblance to its three-dimensional counterpart, the H atom, which has brought it 
the name ``one-dimensional H atom problem" (c) diverse physical applications such as in the exciton in 
high-temperature superconductivity, semiconductors, polymers, 1D electron gas at the helium surface and 
Wigner crystal, etc. Being a function of $|x|$, this is not analytic (e.g., $x \! = \! 0$ is not just a pole) 
and note that the independent variable spans the whole $x$ axis including origin. Acceptable solutions 
must satisfy the wave equation over entire range of $x$. Its exact solution first appeared in 1959 
\cite{loudon59}, where two unusual features were observed for this 1D system. First, the discrete 
bound-state spectrum was found to be degenerate and ground state corresponded to infinite binding 
energy. A later work \cite{davtyan87} employing momentum representation ascribed these rather peculiar 
features to the hidden $O(2)$ symmetry, which was criticized later. Thereafter, This system has 
been studied by a wide range of mathematical methods, such as generalized Laplace transform, Fourier 
transform, quantum phase space representation, momentum space representation, etc. For the past several 
decades, this apparently ``simple" system, initially considered a pedagogical problem, has thrown some intriguing 
controversy among the researchers, such as the degeneracy in ground and excited states, hidden symmetry 
connection, etc., some of which still remain unresolved. In contrast to the 1D H atom potential, the
other 1D Coulomb potential $V(x) \! = \! -Ze^2/x$ has been explored relatively less \cite{reyes99,ran00}.

Quantum mechanical solution of the celebrated 3D Coulomb problem, the H atom, was published as early as in 
1926 by Schr\"{o}dinger in a series of papers by solving an eigenvalue equation for energy of this system. 
Quantum mechanical description of H atom via a central Coulomb potential holds a unique distinction of one 
of the very few realistic physical systems to offer separable and exactly solvable solutions within 
\emph{both} non-relativistic and relativistic picture \cite{semay93}. Innumerable works have been done in the 
following years to obtain valuable insights into this simplest atomic system in 2,3 and N dimensions, from 
both mathematical and physical perspectives \cite{nieto79,cavalli86,morales86,avery92,aquilanti97,aquilanti97a,
nouri99}. Although many textbooks on elementary quantum mechanics present H atom as a closed case, this 
prototypical system continues to offer many new and interesting features, such as its dynamic nature, or its
evolution with time under the influence of a strong electromagnetic field or its  
behavior in a reference frame of arbitrary dimensions. It is now well-known that most of the 3D results 
actually have an N-dimensional counterpart, including Runge-Lenz vector, symmetry properties, Clebsch-Gordan 
coefficients, etc. Symmetry in these higher dimensional systems manifests into the separation of corresponding 
time-independent Schr\"{o}dinger equation into a radial and angular part in hyperspherical coordinates, in 
distinct analogy to the spherical coordinates in 3D and polar coordinates in 2D. In other words, the angular 
solutions emerge as eigenfunctions of a generalized angular momentum operator, usually referred to as 
\emph{hyperspherical harmonics}. Another interesting feature of the N-dimensional Coulomb problem is that its 
solution is connected to the D ($ \! = \! 2N-2$) dimensional harmonic oscillator potential \cite{nouri99}. In other 
words, there exists a transformation which will turn the radial equation for a generalized N-dimensional 
Coulomb potential into that of a D-dimensional harmonic oscillator. This link consists of a map 
$r^2 \! = \! \rho$ and the two are related in following manner,
\begin{equation}
R_{nl}(N;\rho) = \Lambda \Phi_{n',L} (D;r).
\end{equation}
Here, the constant $\Lambda$ arises due to the respective eigenfunctions being normalized in different 
dimensions and is given by,
\begin{equation}
\Lambda= \left( \frac{1}{2[n+\frac{1}{2}(N-3)]^{N+1}} \right)^{1/2}.
\end{equation}
This equation holds only for certain values of $N$ and $D$, and there are also certain relations between the 
respective quantum numbers that must be fulfilled, such as,
\begin{equation}
D=2N-2, \ \ \ \ n'=2(n-1), \ \ \ \ L=2l
\end{equation}

The singular potentials of $1/r^n$, with $n \ge 2$ are of significant current interest. The $n \! = \! 2$ potential 
has relevance in three-body problem in nuclear physics, as well as point-dipole interactions in molecular 
physics \cite{bedaque99,bedaque99a}. Historically, one of the first important difficult cases in dealing 
with highly singular potentials was encountered in the quantum mechanical study of a strongly attractive 
$1/r^2$ term in the Hamiltonian \cite{case50}. This potential shows numerous fascinating features rich in 
physics and mathematics. For example, being uniquely and interestingly placed in the borderline of so-called 
regular and singular potentials, this defines a \emph{transition} point in non-relativistic quantum mechanics 
\cite{frank71}. The $1/r^2$ interaction (in addition to the two-dimensional delta function), is also shown to 
exhibit the phenomenon of anomalous symmetry breaking, wherein a symmetry present in the system at a classical 
level is broken by introducing quantization in to the picture \cite{coon02}. Case $n \! = \! 3$ is used to describe the 
tensor force between nucleons in nuclear physics. Also in the perturbation theory of nuclear interactions 
\cite{kolck99}, proper renormalization of this potential constitutes an important step. Interaction of an atom 
with a flat wall, at short distances, is governed by an attractive van der Waals potential proportional to 
$-1/r^3$, while at larger distances, by the highly retarded Casimir-Polder potential proportional to $-1/r^4$ 
\cite{casimir48}. $n \! = \! 4$ potential also describes the interaction between a charge and an induced dipole 
\cite{vogt54}.
 
\emph{Strictly speaking}, inverse fourth-power potential is the \emph{only true} singular potential (in the 
context of non-relativistic quantum mechanics, these are having a singularity at least as great as the inverse square at 
the center of force \cite{frank71}) and one of the very few besides the afore-mentioned Coulomb ($r^{-1}$) as 
well as harmonic ($r^2$), Morse and P\"{o}schl-Teller potentials, etc., to offer \emph{exact} analytical 
solutions (in terms of Mathieu functions) \cite{frank71,newton82}. $n \! = \! 5$ corresponds to a perturbation 
correction to the tensor force in the nuclear potential \cite{kolck99}. Both $n \! = \! 6$ and 7 are connected to 
the London and Casimir-Polder type van der Waals forces \cite{landau65}. Scattering of atoms by a 
conducting sphere is represented by a $-1/r^6$ potential for small distances and a $-1/r^7$ potential for 
large distances \cite{arnecke07}. Inter-atomic and intermolecular forces at short distances (as strongly 
repulsive due to overlap of the electron clouds) are usually represented by singular 
potentials such as Lennard-Jones (12,6) which has a $\sim 1/r^{12}$ behavior.

The layout of the chapter is as follows. Section II gives an overview of the distinguishing characteristics of
singular and regular potentials. Necessary details of the current methodology is summarized in Section III. Section
IV makes a discussion on results of some central potentials (both singular and non-singular) using this method
with relevance reference to the literature. Finally we end with a few concluding remarks in Section V.

\section{Regular vs. singular}
This article exclusively deals with the non-relativistic quantum mechanical situation, while making only 
casual glances on its relativistic counterpart in some occasions. Since a majority of the potentials 
considered in this work are singular, it would be useful for our future discussion, to differentiate these
from regular potentials.
The motion of a single non-relativistic 
particle in presence of a spherically symmetrical potential $V(|\rvec |) \equiv V(r)$ is governed by the 
following time-independent Schr\"{o}dinger equation in 3D space (henceforth, $\hbar \! = \! m \! = \! 1$ is assumed, 
unless otherwise mentioned);
\begin{equation}
- \frac{1}{2} \nabla^2 \psi(\rvec) + [V(\rvec)-E] \psi(\rvec) = 0
\end{equation} 
Depending on the boundary conditions imposed on such a wave function at large distances, three different 
types of physical situations can be associated with this equation, namely, bound, scattering and 
resonant-state problems. Alternatively, a solution can be prescribed at the boundary conditions at the 
origin. The wave function can be resolved into a sum of products of an $r$-dependent term and an angular 
term. In case of scattering, we thus have the familiar partial-wave expansion,
\begin{equation}
\psi(\rvec)= \sqrt{\frac{k}{\pi}}  \ \frac{1}{4\pi k r} \ \sum_{l=0}^{\infty} (2l+1) \ 
i^l P_l(\cos \theta) \  u_l(r).
\end{equation}
Where $k$ is related to energy $E$ by the expression $k \! = \! \sqrt{\frac{2mE}{\hbar^2}}$ and $P_l(\cos 
\theta)$ signifies the Legendre polynomial. For bound states of angular momentum $l$, only one 
appropriately normalized term would appear from this summation. The corresponding radial wave function 
$u_l(r)$ then satisfies a differential equation of the following form,
\begin{equation}
\left[ \frac{1}{2} \frac{d^2}{dr^2} + {k^2 -\frac{l(l+1)}{r^2}} \right] u_l (k,r)=V(r) u_l(k,r)
\end{equation}
Most of the studies of singular potentials eventually leads to the understanding this radial equation 
and its solutions.

Following \cite{frank71}, a potential $V(r)$ is defined as \emph{regular} at $r \! = \! 0$ if,
\begin{equation}
\lim_{r \rightarrow 0} \ r^2 V(r) = 0
\end{equation}
and \emph{singular} at $r \! = \! 0$ if,
\begin{equation}
\lim_{r \rightarrow 0}\ r^2 V(r) = \pm \infty
\end{equation}
If the limiting values in (6) and (7) are finite, $V(r)$ is termed as \emph{transition} potential. A 
singular potential is classified as repulsive or attractive according to whether the limiting value is, 
respectively, $+\infty$ or $-\infty$. The principal difference between a singular and non-singular 
potential lies in the fact that in the latter case, solutions of the Schr\"odinger equation subject to 
quadratic integrability condition forms a complete orthonormal set, while in the former case, 
solutions are too numerous and hence, over-complete.

In relativistic domain, however, the scenario is much different from the above non-relativistic 
situation. Much weaker infinities in the potential are ``highly singular", so that within a 
relativistic wave equation, even a Coulomb potential is highly singular \cite{case50}. Relativistic 
motion leads to a different singularity criterion; \emph{viz.,} in the vicinity of $r=0$, these 
potentials give an infinite value for the limit,
\begin{equation} 
\lim_{r \rightarrow 0} \ r |V(r)|,
\end{equation} 
in contrast to Eq.~(8) of the non-relativistic case. The regular potentials are defined by a vanishing 
value of this limit, while a finite, non-zero value characterizes a transition potential, which 
includes all those potentials exhibiting Coulomb-like behavior at $r \! = \! 0$. Finally, in contrast to the 
non-relativistic situation, all singular potentials behave like attractively singular in the equations 
of motion, in the relativistic regime.

While the repulsive singular potentials offer no outstanding problems regarding their physical 
interpretation, corresponding attractive case causes serious concern. Physical solutions can be 
determined uniquely for the former, while it is not so for the latter. From a classical point of view, 
a particle moving in an attractive singular potential gives rise to well-defined scattering only if 
certain conditions are fulfilled (such as the impact parameter exceeds certain threshold critical value). 
For any other motions including bounded motions, the particle falls to the origin with an infinite 
velocity. Furthermore, both bound or scattering trajectories are ill-defined unless trajectory tangents 
are matched and also energy as well as angular momentum conservation is maintained at the center of 
force. The scattering problem for an attractive singular potential is also not resolved in quantum picture 
as in the classical case, and is never defined without extraneous physical assumptions (an arbitrary phase 
parameter remains to be assigned). In the bound-state scenario, the attractive singular potentials offer a 
non-unique spectrum consisting of an infinite number of bound states with no lower bounds on the energy. 
For pure power-law potentials, within non-relativistic quantum mechanics, transition from regular to 
singular behavior begins to occur at $r^{-2}$, in much the same way as this happens with the Coulomb 
potential in the relativistic domain. 

\section{The generalized pseudospectral method}
In this article, we are concerned with the accurate bound-state solutions within non-relativistic 
quantum mechanics. As already mentioned, the radial Schr\"odinger equation for a singular potential can be 
solved \emph{exactly} 
only in one case, inverse fourth power, where wave functions are obtained in the form of a modified 
Mathieu functions of generally complex arguments. Thus, approximation methods must be invoked in all other cases. 
A large number of approximate analytic, semi-analytic, numerical methods have been suggested over the past 
several decades. These are appropriately discussed in Section IV. In this section we briefly summarize 
the essential features of our method used in this current work in order to solve the radial eigenvalue 
problem. For a more detailed account on this method, see \cite {roy02,roy02a,roy02b,roy04,roy04a,roy04b,roy05,roy05a,roy05b,
sen06,roy07,roy08,roy08a,roy08b} and the references therein. 

The desired time-independent radial Schr\"odinger equation for a single particle in a non-relativistic 
case can be written as,
\begin{equation}
\hat{H}(r)\ \phi(r) =\varepsilon \ \phi(r).
\end{equation}
The Hamiltonian operator includes usual kinetic and potential energy operators (symbols have their usual 
meanings),
\begin{equation}
\hat{H}(r) =-\frac{1}{2} \ \ \frac{d^2}{dr^2} +v(r),
\end{equation}
with 
\begin{equation}
 v(r) = V(r) + \frac{\ell (\ell+1)}{2r^2} 
\end{equation}
and $V(r)$ is the potential in question. Generally speaking, finite-difference spatial discretization 
schemes often require a large number of grid points to achieve good accuracy presumably because of the fact 
that majority of these methods employ a uniform mesh (non-uniform schemes are used in a few occasions as well, 
e.g., in \cite{killingbeck01}). The generalized pseudospectral (GPS) method, however, can give 
\emph {non-uniform} and optimal spatial discretization accurately, allowing one to work with a denser mesh at 
shorter $r$ regions and a coarser mesh at larger $r$. Additionally, the GPS method is computationally orders 
of magnitude faster than the finite-difference or finite-element methods. 

The principal feature of this scheme lies in approximating a function $f(x)$ defined in the interval 
$x \in [-1,1]$ by a polynomial $f_N(x)$ of order N
\begin{equation}
f(x) \cong f_N(x) = \sum_{j=0}^{N} f(x_j)\ g_j(x),
\end{equation}
such that the approximation is \emph {exact} at \emph {collocation points} $x_j$, i.e.,
\begin{equation}
f_N(x_j) = f(x_j).
\end{equation}
In what follows we employ the Legendre pseudospectral method using $x_0 \! = \! -1$, $x_N \! = \! 1$, where 
$x_j (j=1,\ldots,N-1)$ are obtainable from the roots of first derivatives of Legendre polynomial $P_N(x)$ 
with respect to $x$, i.e., 
\begin{equation}
P'_N(x_j) = 0.
\end{equation}
$g_j(x)$ in Eq.~(13), called cardinal functions, are given by the following expression,
\begin{equation}
g_j(x) = -\frac{1}{N(N+1)P_N(x_j)}\ \  \frac{(1-x^2)\ P'_N(x)}{x-x_j}.
\end{equation}
They have the unique property $g_j(x_{j'}) \! = \! \delta_{j'j}$. Now the semi-infinite domain $r \in [0, \infty]$ 
is mapped onto the finite domain $x \in [-1,1]$ by a transformation $r=r(x)$. One can make use of the 
following algebraic nonlinear mapping,
\begin{equation}
r=r(x)=L\ \ \frac{1+x}{1-x+\alpha},
\end{equation}
where L and $\alpha=2L/r_{max}$ may be termed as the mapping parameters. Furthermore, introducing a relation 
of the form, 
\begin{equation}
\psi(r(x))=\sqrt{r'(x)} f(x)
\end{equation}
in conjunction with a symmetrization procedure \cite{yao93,wang94}, eventually leads to a Hamiltonian in 
the following transformed form as below, 
\begin{equation}
\hat{H}(x)= -\frac{1}{2} \ \frac{1}{r'(x)}\ \frac{d^2}{dx^2} \ \frac{1}{r'(x)}
+ v(r(x))+v_m(x),
\end{equation}
where $v_m(x)$ is given by,
\begin{equation}
v_m(x)=\frac {3(r'')^2-2r'''r'}{8(r')^4}.
\end{equation}
The advantage is that this leads to a \emph {symmetric} matrix eigenvalue problem which can be readily 
solved to give accurate eigenvalues and eigenfunctions. For the particular transformation used in this work, 
$v_m(x) \! = \! 0$. This discretization then leads to following set of coupled equations, 
\begin{widetext}
\begin{equation}
\sum_{j=0}^N \left[ -\frac{1}{2} D^{(2)}_{j'j} + \delta_{j'j} \ v(r(x_j))
+\delta_{j'j}\ v_m(r(x_j))\right] A_j = EA_{j'},\ \ \ \ j=1,\ldots,N-1,
\end{equation}
\end{widetext}
where
\begin{equation}
A_j  = \left[ r'(x_j)\right]^{1/2} \psi(r(x_j))\ \left[ P_N(x_j)\right]^{-1}.
\end{equation}
and the symmetrized second derivative of the cardinal function, $D^{(2)}_{j'j}$ is given by,
\begin{equation}
D^{(2)}_{j'j} =  \left[r'(x_{j'}) \right]^{-1} d^{(2)}_{j'j} 
\left[r'(x_j)\right]^{-1}, 
\end{equation}
with
\begin{eqnarray}
d^{(2)}_{j',j} & = & \frac{1}{r'(x)} \ \frac{(N+1)(N+2)} {6(1-x_j)^2} \ 
\frac{1}{r'(x)}, \ \ \ j=j', \nonumber \\
 & & \nonumber \\
& = & \frac{1}{r'(x_{j'})} \ \ \frac{1}{(x_j-x_{j'})^2} \ \frac{1}{r'(x_j)}, 
\ \ \ j\neq j'.
\end{eqnarray}
It is worth mentioning here that GPS method offers both simplicity of direct finite-difference and/or
finite-element method, as well as the fast convergence of finite basis set method. In this sense, one can have the
cake and eat it too! It has been rigorously proved 
\cite{gottlieb84,canuto88} that the method guarantees an \emph{exponential} (or alternatively termed, 
\emph{infinite-order}) convergence for a problem of a \emph{smooth} or \emph{infinitely 
differentiable} solution (which is often the case) provided that the orthogonal functions of a common 
singular Sturm-Liouville problem are used. Here the `exponential convergence' implies that the error 
of approximate solution decreases asymptotically faster than the algebraic decay of any order. 
Moreover a GPS scheme having (N+1) grid points is usually equivalent in accuracy to a corresponding 
basis-set expansion method involving N basis functions. Many other details pertaining to 
this approach could be found in the references \cite {roy02,roy02a,roy02b,roy04,roy04a,roy04b,roy05,
roy05a,roy05b,sen06,roy07,roy08,roy08a,roy08b}.

Finally, a series of test calculations were done with a variety of potentials in the literature for which
exact/near-exact solutions have been reported, in 
order to optimize its performance with respect to the mapping parameters. In this way, the following 
parameter set ($r_{max} \! = \! 200, \alpha \! = \! 25, N \! = \! 300$) has been consistently used throughout 
this work (unless otherwise mentioned), which seemed to be quite satisfactory for the purpose at hand. 

\section{results and discussion}
This section presents a discussion on the results obtained by using GPS. As will be evident soon, this has 
been very successfully applied to a large variety of physical/chemical problems including static and dynamic
situations. Here we select a cross-section of the most compelling and testing cases, while the rest of these
could be found by inquisitive reader in the references \cite {roy02,roy02a,roy02b,roy04,roy04a,roy04b,roy05,
roy05a,roy05b, sen06,roy07,roy08,roy08a,roy08b}. However, before we move on to 
the various central potentials representing different physical systems, first we give a brief snapshot 
of this approach for one particular potential. For this illustrative purpose, we have chosen the 3D quartic oscillator as a 
prototypical case.  Quartic oscillators have found many notable applications. For example, two- and three-dimensional 
anharmonic oscillators have drawn much interest in far 
infra-red and microwave regions. Four-membered ring molecules (such as trimethylene oxide, cyclobutanone 
etc.) are known to have out-of-plane vibrational modes which are predominantly quartic, five-membered 
ring compounds also have ring puckering modes with a significant quartic contribution to the potential. 
Therefore the pure quartic oscillator, the mixed quartic-harmonic oscillators and in general, the 
anharmonic oscillators have been subject to intense study in both quantum field theory and chemical 
physics for a long period of time and the interest still continues to grow \cite{bell70,mathews82,
shanley88a,shanley88b,lakshmanan94,skala97,delabaere97}. General analytical solutions of the quartic 
oscillator are unknown; although some special cases \cite{skala97} are reported for only certain states 
of a 1D quartic oscillator where solutions could be found analytically (e.g., the modified quartic oscillator where 
the potential depends on $|x|$). Thus there has been considerable interest to study the bound-state 
spectra of these physically as well as chemically important systems using a wide variety of mathematical 
methods, such as WKB method, phase-integral approach etc. 

\begingroup
\squeezetable
\begin{table}  
\caption {\label{tab:table1} Comparison of pure 3D  quartic oscillator energies for some high-lying states; 
$\nu=48,49$ and $l=0-9.$ Taken from ref.~\cite{roy05a}.}
\begin{ruledtabular}
\begin{center}             
\begin{tabular} {cclllll}
{$\nu$} & $l$ & GPS \cite{roy05a} & Variational \cite{bell70}  & Analytical \cite{mathews82} & 
Finite-difference \cite{killingbeck82} & Asymptotic shooting \cite{holubec90} \\
\hline
48 & 0 & 250.183358697 & 250.183351 & 250.183369 & 250.183359  & 250.1833586971 \\
49 & 1 & 256.916238928 & 256.916220 & 256.916238 & 256.916239  & 256.9162389286 \\
48 & 2 & 250.096690608 & 250.096679 & 250.096671 & 250.096691  & 250.0966906080 \\
48 & 8 & 249.144812457 & 249.144801 & 249.1452   &             & 249.1448124575 \\
49 & 9 & 255.664161642 & 255.664146 & 255.66480  &             & 255.6641616427 \\
\end{tabular}
\end{center}
\end{ruledtabular}
\end{table}
\endgroup

Each energy level in a 3D isotropic harmonic oscillator is $\frac{1}{2} (\nu+1) (\nu+2)$-fold degenerate. 
This degeneracy is associated with the two angular momentum quantum numbers $l,m$. Each $l$th level is 
$(2l+1)$-fold degenerate, corresponding to the $(2l+1)$ linearly independent states with $m=l, l-1, \cdots, 
0, \cdots, -l$. Upon introduction of a radial perturbation to a harmonic oscillator (such as a pure 3D 
quartic oscillator) the degeneracy present in a harmonic oscillator is partly removed. For any given 
$\nu$, levels with different values of $l$ are split, but the $m$-type degeneracy of each level remains. 
In other words, $l$ remains a good quantum number; each $l$th level being $(2l+1)$-fold degenerate. Only 
an angular perturbation removes the degeneracy of these $m$ levels. Table I compares some specimen results 
for odd- and even-parity high-lying excited states of the 3D pure quartic oscillator. Vibrational quantum 
numbers correspond to $\nu \! = \! 48,49$, and angular momentum quantum number is varied from $l \! = \! 0-9$. Note that 
these two quantum numbers must match in parity. The present results are quoted from \cite{roy05a}. It is 
worth mentioning that while numerous works have been presented for low-lying states (especially the ground 
states), similar successful attempts for accurate treatment of high-lying states such as those concerned 
here, have been dramatically less, because of the difficulties faced. Our results are compared with some of 
the carefully selected best literature values, \emph{viz.}, (a) linear variational calculations involving 
diagonalization of large order matrices ($800 \times 800$) \cite{bell70}, (b) approximate analytical 
formulas derived from a scaled oscillator approach \cite{mathews82} (c) finite-difference numerical 
result \cite{killingbeck82} and the (d) asymptotic shooting method \cite{holubec90}. Clearly, amongst 
these literature values, \cite{holubec90} appears to be the most accurate, where solutions are expressed 
in terms of finite polynomials, requiring straightforward determination (integration) of zeros of such 
polynomials. As evident, our present results for all these states match exactly up to the 9th decimal 
place with those of \cite{holubec90}, which manifestly demonstrates the power and efficacy of this method 
for higher excitations. Similar kind of tests have been made for other cases such as Morse oscillator, 
an anharmonic oscillator with a quartic perturbation \cite{roy04a}, charged harmonic oscillator \cite{roy04}, 
Hulth\'en potential \cite{roy05}, harmonic potential including an inverse quartic and sextic perturbation as 
well as a Coulomb potential with a linear and quadratic coupling \cite{roy05a}, etc., which offer conditionally
exact solution for certain states. In all such cases, near-exact solutions were obtained from GPS method. 
Now we proceed for the discussion on individual potentials.

\subsection{Spiked Harmonic Oscillator}
A class of singular potentials defined by the following Hamiltonian,
\begin{equation}
\mathrm{H}=\mathrm{p}^2+\mathrm{r}^2+\lambda \mathrm{r}^{-\alpha} 
\equiv \mathrm{H_0}+\lambda \mathrm{r}^{-\alpha}, \ \ 
\mathrm{r} \in [0, \infty], \ \ \alpha > 0
\end{equation}
where $\mathrm{p}=-i\ \partial /\partial \mathrm{r}$, has been termed as the spiked harmonic oscillator 
(SHO). $\mathrm{H_0}$ stands for the simple harmonic oscillator Hamiltonian, whereas the coupling parameter
$\lambda$ and positive constant $\alpha$ determine \emph{strength} of the perturbative potential and 
\emph{type} of singularity at the origin respectively. Ever since the fascinating work of \cite{harrell77} on its 
ground-state energies in the mid 70s, an enormous amount of works have been published for its studies
\cite{klauder73,defacio74,ezawa75,detwiler75,klauder78,killingbeck82,znojil89,znojil90,navarro90,navarro91,
kaushal91,fernandez91,guardiola92,navarro92,kaushal92,znojil92,torres92,breton93,landtman93,varshni93,
navarro94,miller94,hall95,buendia95,handy96,bhattacharyya96,trost97,hall98,hall98a,mustafa99,mustafa00,
hall00,hall00a,killingbeck01,hall01, chakrabarti02,chakrabarti02a,hall02,guardiola02,saad03,saad03a,
bhattacharjee03,roy04,bandyopadhyay05,bandyopadhyay05a,roy08b}. 
This is not only due to its widespread applications in atomic, molecular, nuclear and particle physics 
but also because of multitude of inherent interesting properties from mathematical physics point of view. This 
gives rise to an interesting situation recognized long times ago, i.e., no dominance of either of two
terms in the interaction potential for extreme values of $\lambda$. In other words, one never deals with 
\emph{small} perturbations \cite{navarro90,guardiola92}. For all $\lambda \! \! \rightarrow \! \!0$, 
$\lambda \mathrm{r}^{-\alpha}$ adds an infinite repulsive barrier near the origin. On the other hand, in 
the limit of $\lambda \!\!\rightarrow \! \! \infty,$ one can not ignore the $r^2$ term. In fact, the 
harmonic term can never be neglected, for it is needed for the existence of definite ground states 
\cite{case50,spector64}. Thus the potential resembles a wide valley extending to $\infty$ \cite{navarro91}.
Another distinctive feature related to this potential is that they exhibit the so-called \emph{Klauder 
phenomena}, \emph{viz.}, once the perturbation is turned on, complete turn-off is impossible; permanent
irreversible vestigial effects of the interaction persists \cite{klauder73,defacio74,ezawa75,klauder78}. 
In particular, a sufficiently singular potential can not be smoothly turned off ($\lambda \! \rightarrow 
\! 0$) in the Hamiltonian ($H \! = \! H_0+\lambda V$) to restore the free Hamiltonian $H_0$. It has been shown 
that \cite{harrell77} for an SHO, the familiar Rayleigh-Schr\"odinger perturbation diverges according to the relation 
$n \! \ge \! \frac{1}{\alpha -2}$, where $n$ is the order of perturbation term. Accordingly, the first-order perturbation 
correction diverges for $\alpha \ge 2$, second order for $\alpha \ge 5/2$, etc. This potential also exhibits
the \emph{super-singularity} phenomenon \cite{detwiler75} in the region of $\alpha \ge 5/2$, i.e., every 
matrix element of the potential becomes infinite. They have been discussed in the interesting context of
degeneracy in one-dimensional systems \cite{bhattacharyya96}. In another work \cite{bhattacharjee03}, it
was found that, the perturbation theory in $\lambda$ has an ultraviolet divergence for a range of $\alpha$,
which causes the perturbation series to be ordered by $\lambda^Z$; $Z$ being a fraction less than unity.

At this stage it would be convenient for our purpose to discuss a simpler special case of $\alpha \! = \! 1$, 
the so-called \emph{charged harmonic oscillator}, before we proceed for the general stronger spikes 
($\alpha \ne 1$). This is a non-supersingular spiked oscillator characterized by a perturbative term of
the Coulomb form $\lambda/r$, and has been studied in considerable detail \cite{navarro92}. It is possible 
to identify three distinct regions depending on the value of \emph{effective} coupling constant, \emph{viz.},
(a) \emph{Coulomb} region corresponding to large negative values of $\lambda$ (b) \emph{strong-coupling} 
region corresponding to large positive values of $\lambda$ (c) \emph{weak-coupling} region having small 
(positive or negative) values of $\lambda$. For (a), (b) approximate Rayleigh-Ritz perurbative expansions 
for ground (and some excited) state energies were developed by means of a combination of hypervirial 
relation and Hellmann-Feynman theorem. In the Coulomb case, this is given as, 
\begin{equation}
E(\lambda)= \lambda^2 \left[-\frac{1}{4}+\frac{12}{\lambda^4}-\frac{1032}{\lambda^{8}}+
\frac{348864}{\lambda^{12}}-\frac{211519200}{\lambda^{16}}+
\frac{188054861568}{\lambda^{20}}+ \cdots \right]
\end{equation}
whereas for the strong-coupling case this becomes ($\mu=(2/\lambda)^{1/3})$, 
\begin{equation}
E= 3\mu^{-2}+\sqrt{3}+\frac{7\mu^2}{36}+\frac{37\mu^4}{432\sqrt{3}} + \frac{2573\mu^6}{139968} 
+ \frac{168233\mu^8}{2239488\sqrt{27}} +\cdots
\end{equation}

It is also known that there could be an indirect, somewhat involved path to connect the strong-coupling 
regime ($\lambda \! \! \rightarrow \! \! \infty$) with the small-coupling regime, as well as to connect
Coulomb regime ($\lambda \! \! \rightarrow \! \! -\infty$) with the small-coupling regime. However, as 
yet no direct link has been found to connect the $+\infty$ and $-\infty$ regions. Another very important 
characteristic of the charged oscillator is that this offers an infinite set of \emph{elementary solutions}
\cite{navarro92} for certain selected values of positive coupling constant only. These solutions are 
typically of the form of a polynomial multiplied by a Gaussian and are possible for both ground as well excited 
states. Some of these, taken from \cite{roy04}, are displayed in Table II. Note that ``Exact'' results 
have been divided by a 2 factor for consistency with the literature. We see excellent agreement of current 
method with ``Exact'' results for all values of $\lambda$. Ground states for general $\lambda$ (both positive 
and negative) have been reported in \cite{navarro92,roy04}, whereas first three excited states corresponding 
to $l=0,1,2,3$ have been studied in \cite{roy04}. This concludes our discussion on charged oscillator; more 
detail could be found in \cite{navarro92,roy04}.

\begingroup
\squeezetable
\begin{table}   
\caption {\label{tab:table2}Some elementary solutions (in a.u.) of the charged harmonic oscillator 
($\alpha=1$). Ground state energies are given for different values of $\lambda$. Results from GPS method 
are taken from \cite{roy04} and ``Exact'' results are quoted from \cite{navarro92}.}
\begin{ruledtabular}
\begin{center}
\begin{tabular}{cccccc}
$\lambda$ & E(GPS) &  E(Exact)   &  $\lambda$   &   E(GPS)  &  E(Exact)  \\  
\hline
0                       & 1.49999999999  &  1.5   & 2 & 2.49999999999  &  2.5   \\
$\sqrt{20} $            & 3.50000000000  &  3.5   & $(30+6\sqrt{17})^{1/2}$   & 4.49999999999  &  4.5   \\
$(70+6\sqrt{57})^{1/2}$ & 5.49999999999  &  5.5   & 14.450001026966         & 6.49999999999  &  6.5   \\
\end{tabular}
\end{center}
\end{ruledtabular}
\end{table}
\endgroup

Let us now return back to SHO. 
From a variational analysis \cite{detwiler75}, the eigenvalues of SHO for $2 \le \alpha \le 3$ were given
by asymptotic series to first order for positive $\lambda$. But for $\alpha \ge 3$, the ground-state 
eigenvalues are given by,
\begin{equation}
E_0(\lambda)=3 + k \lambda^{\nu} + O (\lambda^{\nu})
\end{equation}
whereas for $\alpha=3$, these are obtained as,
\begin{equation}
E_0(\lambda)=3+k'\lambda \log (\lambda) + O(\lambda)
\end{equation}
where $k,k'$ are to be determined variationally. Later, a modified Rayleigh-Schr\"odinger series was put
forth \cite{harrell77} by exploiting the standard WKB approximation for lowest few orders. This success encouraged
other workers to further develop a special perturbation theory, known as \emph{singular perturbation 
theory} to obtain first few terms of perturbed $\lambda$-expansion for different values of $\alpha$. To
this end, asymptotic series for ground-state eigenvalues of the SHO Hamiltonian are explicitly written as,
\begin{eqnarray}
E_0(\alpha \ge 4,\lambda) & = & 3+\frac{4\nu^{2\nu} \Gamma(1-\nu)}{\sqrt{\pi}\Gamma(1+\nu)} \lambda^{\nu} 
                        + O(\lambda^{2\nu})  \nonumber  \\
E_0(3<\alpha<4,\lambda) & = & 3+\frac{4\nu^{2\nu} \Gamma(1-\nu)}{\sqrt{\pi}\Gamma(1+\nu)} \lambda^{\nu} 
 -\frac{4 \nu \Gamma(\frac{3-\frac{1}{\nu}}{2})}{(1-\nu) \sqrt{\pi}} \lambda + O(\lambda^{2\nu}) \nonumber \\
E_0(\alpha=3,\lambda) & = & 3-\frac{4}{\sqrt{\pi}} \lambda \log (\lambda) - \frac{10c}{\sqrt{\pi}} \lambda
                        + O(\lambda^{2} \log^2 (\lambda)) \nonumber \\
E_0(5/2<\alpha<3,\lambda) & = &3+\frac{4\nu^{2\nu} \Gamma (1-\nu)}{\sqrt{\pi}\Gamma (1+\nu)} \lambda^{\nu}
  +\frac{2\Gamma (\frac{3-\alpha}{2})}{\sqrt{\pi}} \lambda + O(\lambda^{2\nu})  
\end{eqnarray}
Here $\nu=1/(\alpha-2)$ and $c=0.5772156649$ is the Euler's constant. It is worth noting that $\alpha \! = \! 5/2$
bears a close analogy with the low-density expansion of energy of a many-body boson system at zero 
temperature \cite{baker71}, \emph{viz.},
\begin{equation}
E/N=(2\pi\hbar^2/m) \rho a[1+C_1(\rho a^3)^{1/2} + C_2 \rho a^3 \ \mathrm{ln} (\rho a^3) + 
     C_3 \rho a^3 + \cdots ]
\end{equation}

In another development \cite{navarro90}, ground-state energy was obtained by making use of a functional 
space spanned by the solution of Schr\"odinger equation for a linear harmonic oscillator within a 
variational framework, followed by standard diagonalization of symmetric matrices. A strong coupling 
expansion ($\alpha \ge 2$) for positive $\lambda$ was derived for approximate ground-state energy. For
example, for $\lambda=5/2$, this reads as,      
\begin{equation}
E(\lambda)= \frac{9}{5} \left( \frac{5}{9} \right)^{4/9} \lambda^{4/9} + 
\left( \frac{9}{2} \right)^{1/2} + \cdots 
\end{equation}
For same $\alpha$, the fourth-order large coupling perturbation expression is,
\begin{equation}
E(\lambda)= \frac{9}{5} \left( \frac{5\lambda}{4} \right)^{4/9} + \left( \frac{9}{2} \right)^{1/2} 
+\frac{77}{288} \left( \frac{4}{5\lambda} \right)^{4/9} - 
\frac{1967}{27648} \left( \frac{2}{9} \right)^{1/2} \left( \frac{4}{5\lambda} \right)^{8/9} + \cdots
\end{equation} 

The weak coupling expression using perturbation theory up to second order as well as the strong 
coupling expansions up to 10th order (in algebraic form) were obtained for the non-singular SHO 
($\alpha < 5/2$) through a re-summation technique \cite{navarro91}. In the above work, also an attempt was 
made to find a path connecting ground-state energy in the weak coupling regime with that of the strong 
coupling regime. Similar weak-coupling expansions for ground-state energies were given for the special
case of $\alpha \! = \! 2$ in \cite{breton93}. Ground-state energy of the particular case $\alpha \! = \! 4$ was studied
by using a non-orthonormal basis set satisfying the correct boundary conditions \cite{navarro94}. Through
a logarithmic perturbation theory \cite{bandyopadhyay05}, closed-form expressions for energy and wave 
function correction terms were obtained for ground states.

In another analytical method, namely, pseudo-perturbative shifted-$l$ expansion technique (PSLET) 
\cite{mustafa99}, it was possible to obtain both eigenvalues and eigenfunctions in one batch. This uses
$1/\bar{l}$ as a perturbation expansion parameter, where $\bar{l} \! = \! l-\beta$ ($l$ is a quantum number and 
$\beta$ is a suitable shift introduced to avoid the trivial case $l \! = \! 0$). This has been used to study bound states of
D-dimensional spiked harmonic oscillator spectra \cite{mustafa00} as well, taking into account the effect of
inter-dimensional degeneracies arising out of the isomerism between angular momentum and dimensionality of 
the central force Schr\"odinger equation. Other approximate variational methods have also been attempted
\cite{fernandez91}. These states have been studied by a modified WKB approximation \cite{trost97}. Upper
and lower bounds of ground as well as excited states have been developed \cite{hall98} by means of envelope 
theory through some convenient smooth transformation. A rigorous variational method suitable for the complete
set of discrete energy eigenvalues has been established \cite{hall02}, which also remains valid for arbitrary
angular momentum quantum number, in general N-dimensional case. This was done by expressing the SHO Hamiltonian
as a perturbation of the singular Gol'dman and Krivchenkov Hamiltonian $H_0$. Further it was proved that 
zeroth order eigenfunctions generated by $H_0$ form a suitable singularity-adapted basis for the appropriate 
Hilbert space of the full problem. Ground-state bounds have been investigated by means of potential 
envelopes \cite{hall95}. An eigenvalue moment method has been proposed for accurate estimation of ground states
of singular potentials \cite{handy96}. 

The SHO problem was numerically solved through finite-difference scheme \cite{killingbeck82,killingbeck01},
a Lanczos grid method \cite{torres92}. In the latter, Hamiltonian was discretized on a grid with a 10th
order finite-difference formula for kinetic energy and a starting function $r^8e^{-r^2}$. Applications were
made for $s$ states for several values of $\alpha, \lambda$. Accurate numerical solutions were found 
\cite{buendia95} by modifying the usual analytic continuation method of \cite{holubec85,holubec90}, which
was initially applicable to potentials whose solutions do not have essential singularities. This was possible
by introducing a family of non-singular potentials which depend on a parameter $R_c$ and which approach the 
exact potential as $R_c$ tends to zero. A non-perturbative but completely convergent algorithm \cite{miller94}, 
and formally identical to the Lanczos method was proposed for ground states. An accurate algorithm was presented
where a discretized symmetric expression was derived for the transformed Schr\"odinger equation 
\cite{guardiola92}, which could be efficiently solved by standard routines for tridiagonal matrices efficiently.
This involved a coordinate transformation either of the form $r=Kx/(1-x)$ (parametrized Euler transformation) or
$1+Kr= e^{Kx}$ \cite{killingbeck01}, where the adjustable parameter $K$ needs to be chosen judiciously.

In spite of all the above attempts, a general method which can offer accurate reliable results for both these 
potential parameters for arbitrary states (covering both ground as well as various excited states, especially 
the higher ones), have been
very rare. Thus, for example, physically meaningful and high accuracy results are obtainable by only a few of 
the methods discussed above. Moreover, some of these methods provide high-quality results for certain
type of parameters, while performing rather poorly for other sets. An enormous amount of work has been reported
for ground states; excited states have received much less attention presumably due to the inherent difficulties 
encountered with them. Excepting a very few rare studies (for example, \cite{hall01,roy04}), these have almost 
exclusively dealt with eigenvalue determination; nature of eigenfunctions has been explored only in some rare
occasions. Last, but not the least, some of these methods are often fraught with rather tedious, cumbersome  
mathematical complexities. The GPS method provides a simple general and easily affordable scheme where all these
above mentioned discomfitures are either completely removed or partly alleviated by invoking an \emph{optimal,
non-uniform} effective spatial grid. General applicability of the approach is amply demonstrated in Tables III,
IV for some representative parameter sets of the potential for arbitrary states. Literature results are
quoted, for comparison, wherever possible. Clearly very accurate results are obtained for a large range of 
coupling parameters for low as well as high states. More detailed results on energies, expectation values, radial
densities through the GPS method are available in \cite{roy04}.

\begingroup
\squeezetable
\begin{table}  
\caption {\label{tab:table3} Comparison of ground-state energies E (in a.u.) of the SHO
with $\alpha=4$ and 6 for selected values of $\lambda.$ See \cite{roy04} for details.} 
\begin{ruledtabular}
\begin{center}
\begin{tabular}{cllll}
$\lambda$ & \multicolumn{2}{c}{Energy ($\alpha=4$)} & 
\multicolumn{2}{c}{Energy ($\alpha=6$)} \\ 
\cline{2-3} \cline{4-5}
      & GPS\cite{roy04}     & Literature   & GPS\cite{roy04}      & Literature \\   \hline
0.001 & 1.53438158545 & 1.53438158545$^{\mathrm{a}}$, 1.534385$^{\mathrm{b}}$ & 
1.63992791296 & 1.63992791296$^{\mathrm{a}}$  \\
1     & 2.24708899168 & 2.24708899168$^{\mathrm{a}}$,
     2.24709$^{\mathrm{b,c}}$  & 2.32996998478 & 
2.32996998478$^{\mathrm{a}}$,2.329970$^{\mathrm{c}}$  \\ 
   &  & 3.3033112560$^{\mathrm{d}}$   &    &  3.00160451$^{\mathrm{d}}$  \\
1000  & 10.6847312660 & 10.6847312660$^{\mathrm{a}}$, 10.68473$^{\mathrm{b}}$, & 
6.35930853290 &    \\
   &   &  10.684731265$^{\mathrm{d}}$   &    &    \\
\end{tabular}
\begin{tabbing}
$^{\mathrm{a}}\!$ Ref.~\cite{buendia95}. \hspace{0.3in} \= $^{\mathrm{b}} \!$ Ref.~\cite{navarro94}. 
\hspace{0.3in} \= $^{\mathrm{c}} \!$ Ref.~\cite{fernandez91}. \hspace{0.3in}  \= 
$^{\mathrm{d}} \!$ Ref.~\cite{handy96}.
\end{tabbing}
\end{center}
\vspace{-0.2in}
\end{ruledtabular}
\end{table}
\endgroup

Now we discuss a slightly modified Hamiltonian from that defined in Eq.~(25), \emph{viz.},
\begin{equation}
H=-\frac{d^2}{dr^2} + ar^2 +\frac{b}{r^4}+\frac{c}{r^6},  \ \ \ \ a,c>0,  r \in [0,\infty] 
\end{equation}
which is a non-trivial generalization of the SHO Hamiltonian and has also been studied in considerable detail in 
the literature. For example, in \cite{guardiola92}, a large-order strong coupling expansion, in terms of
some \emph{ad hoc} expansion parameter was proposed for large anharmonicity constants corresponding to 
either of b,c large or both of them large. For small anharmonicity constants, lowest order correction to
the perturbed energies were also obtained. 

A very important feature of this central singular 
potential is that it provides conditionally exact solutions for certain values of the potential parameters. 
From very early stage of the development of quantum mechanics, there is continual interest in solving
Schr\"odinger equation exactly. However, as we know, exact solutions could 
be found for very few potentials. For systems with one degree of freedom, supersymmetric quantum mechanics 
together with shape invariance has been found to be one of the most successful techniques for understanding
the exact solvability. Of late, due attention has been paid to different class of potentials which are 
quasi-exactly solvable (QES) and conditionally exact solvable (CES). In the former case, only a finite 
number of eigenstates can be found exactly. A more interesting case is offered by the CES potential, which
is intermediate between the exact solvable and QES potential. For, exact eigenvalues are obtainable 
only when potential parameters satisfy certain conditions. The wave function ansatz technique is one very
common procedure used for this purpose, which is purely mathematical and usually fails to resolve the 
physical reason for conditional solvability. However, this technique is not straightforward for 
arbitrary states. Recently, super-potential ansatz technique has also been proposed.
Bound states in this potential for $c=0$ 
\cite{znojil89} and $c \ne 0$ \cite{znojil90} have been investigated nearly two decades ago. A simplified
ansatz for eigenfunctions has offered exact ground-state energy in the following closed form 
\cite{kaushal91},
\begin{equation}
\phi_0(r)=N_0 \ r^{(3+b/\sqrt{c})/2} \ e^{-\frac{1}{2}(\sqrt{a}r^2 + \sqrt{c} r^{-2})}
\end{equation}
with the ground-state energy given as, 
\begin{equation}
E_0=\sqrt{a}(4+b/\sqrt{c})
\end{equation}
However, for this to be satisfied, the potential parameters must be related as follows:
\begin{equation}
(2\sqrt{c}+b)^2  = c[(2l+1)^2+8\sqrt{ac}]
\end{equation}

\begingroup
\squeezetable
\begin{table}  
\caption {\label{tab:table4}Comparison of energies of $\ell \ne 0$ states (in a.u.) of the 
SHO with $\alpha=4$ (top) and 6 (bottom) for selected $\lambda$s. Un-superscripted GPS
results are quoted from \cite{roy04}.} 
\begin{ruledtabular}
\begin{center}
\begin{tabular}{llll}
$\ell$ & $\lambda=0.001$ &  $\lambda=0.1$ & $\lambda=10$ \\ \hline
 5 & 6.50002020182(6.50002020182)\footnotemark[1] & 6.50201821626(6.50201821626)\footnotemark[1] & 6.68566506197  \\ 
20 & 21.5000012507 & 21.5001250765 & 21.5124915772  \\
50 & 51.5000001999 & 51.5000200019 &  51.5020000374  \\
\hline 
5  & 6.50000577192 & 6.50057643602 & 6.55258902874  \\ 
20 & 21.5000000675 & 21.5000067609 & 21.5006759799  \\
50 & 51.5000000040 & 51.5000004123 & 51.5000412410  \\
\end{tabular}
\end{center}
\end{ruledtabular}
\footnotetext[1] {Ref.~\cite{saad03a}.}
\end{table}
\endgroup

Later, a similar approach was employed for first excited states \cite{kaushal92}. The harmonic 
potential with an inverse quartic and sextic anharmonicity was solved numerically \cite{landtman93} 
for some lowest states using B-spline basis sets. Continuing along the same line, four sets of solutions 
were obtained, including one constraint equation for each set \cite{varshni93}; 
furthermore it was found that the analytical expression for energy agrees with numerical result for any
one among the ground, first and second excited states, depending on the particular constraint condition.
Conditional exact solutions were studied by \cite{chakrabarti02,chakrabarti02a} in the light of
supersymmetric quantum mechanics and shape invariance. A Hill determinant approach was attempted
\cite{znojil92}.

In parallel to above works, recently there has been considerable interest in studying the so-called
generalized spiked harmonic oscillator(GSHO), defined as, 
\begin{equation}
H=-\frac{d^2}{dr^2} +v(r); \ \ \ v(r)=r^2+\frac{A}{r^2}+\frac{\lambda}{r^{\alpha}}, \ A \ge 0
\end{equation}
Here $\lambda, \alpha$ are two real parameters and obviously SHO is a special case of GSHO with $A \! = \! 0$. Both 
variation and perturbation methods were employed for their studies \cite{hall98a,hall00,hall00a,hall01,saad03,saad03a}. 
Using the exact Gol'dman-Krivchenkov wave functions
(which constitute an orthonormal basis for the Hilbert space $L^2(0,\infty)$), compact closed form expressions
have been derived \cite{hall98a,hall01} for the required singular-potential integrals (or matrix elements) of the 
form $\langle m|x^{-\alpha}|n \rangle$. Variational bounds \cite{hall00} have been examined on the light of 
above as well. A first-order perturbation series was developed \cite{hall01} for wave functions in terms of 
generalized hyper-geometric functions. The modified perturbation theory of \cite{harrell77} was extended to 
obtain perturbation expansions for the GSHO problem \cite{saad03}. These expressions remain valid for small 
values of coupling $\lambda \! >\! 0$ and corroborate the results of those obtained for SHO \cite{harrell77}.
In \cite{saad03a}, using a perturbation expansion up to 3rd order, lower and upper bounds of the GSHO ground
states for small coupling parameter $\lambda$, were estimated. Weak-coupling perturbation expansions for ground
states were derived in \cite{hall00a}.

\begin{figure}
\centering
\begin{minipage}[t]{0.27\textwidth}\centering
\includegraphics[scale=0.32]{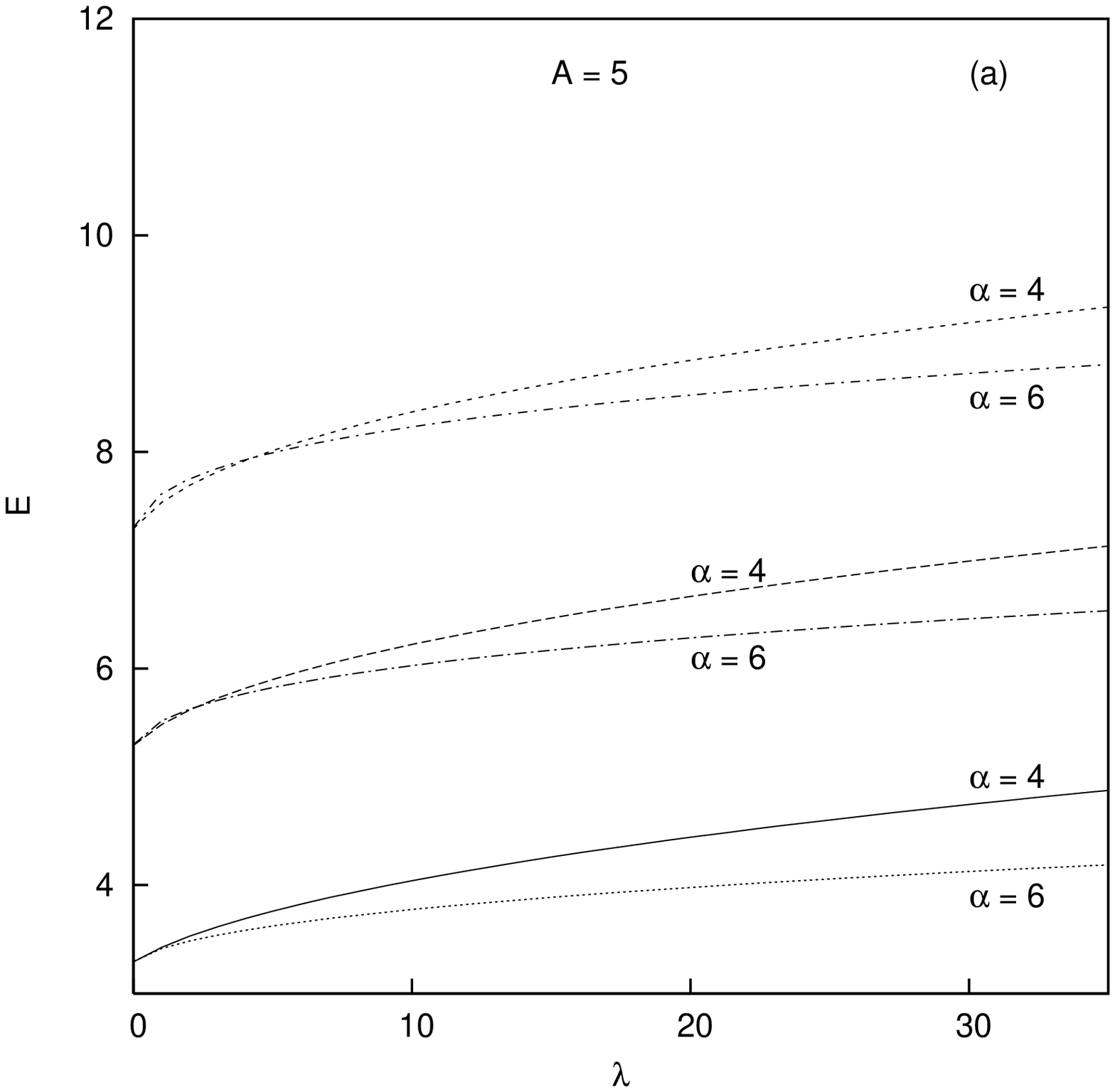}
\end{minipage}
\hspace{0.40in}
\begin{minipage}[t]{0.27\textwidth}\centering
\includegraphics[scale=0.32]{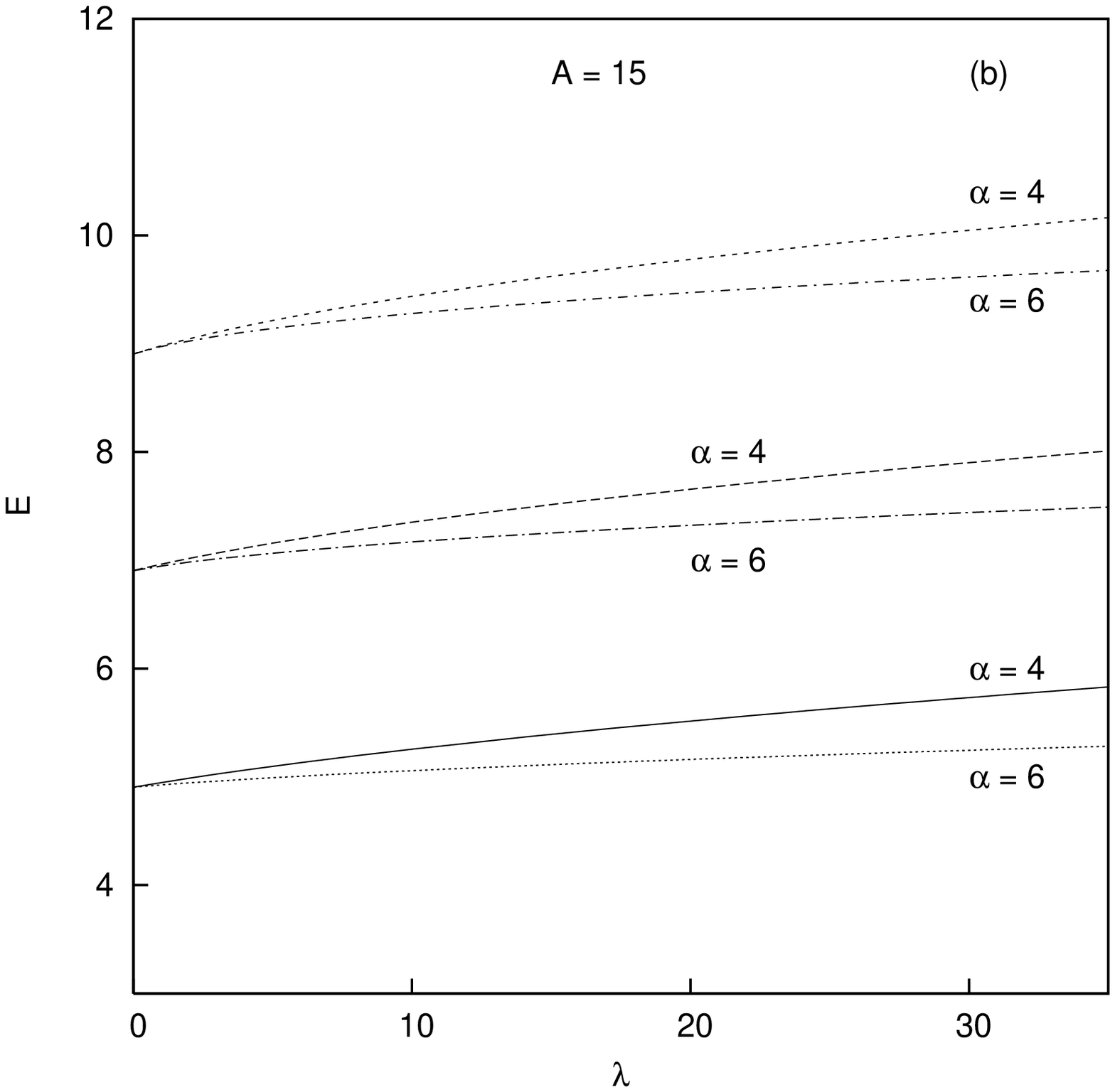}
\end{minipage}
\hspace{0.40in}
\begin{minipage}[t]{0.27\textwidth}\centering
\includegraphics[scale=0.32]{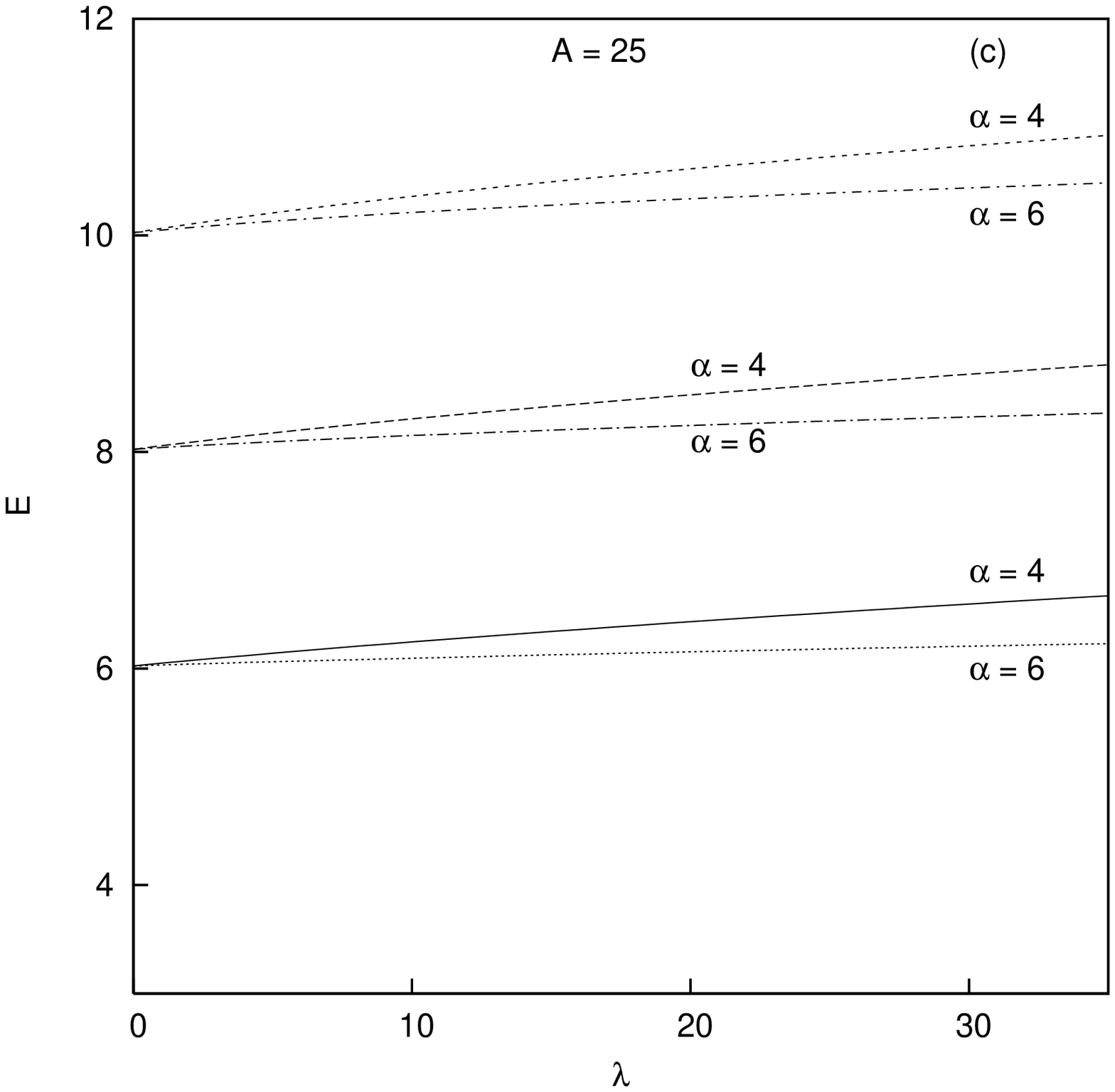}
\end{minipage}
\\[30pt]
\begin{minipage}[b]{0.27\textwidth}\centering
\includegraphics[scale=0.32]{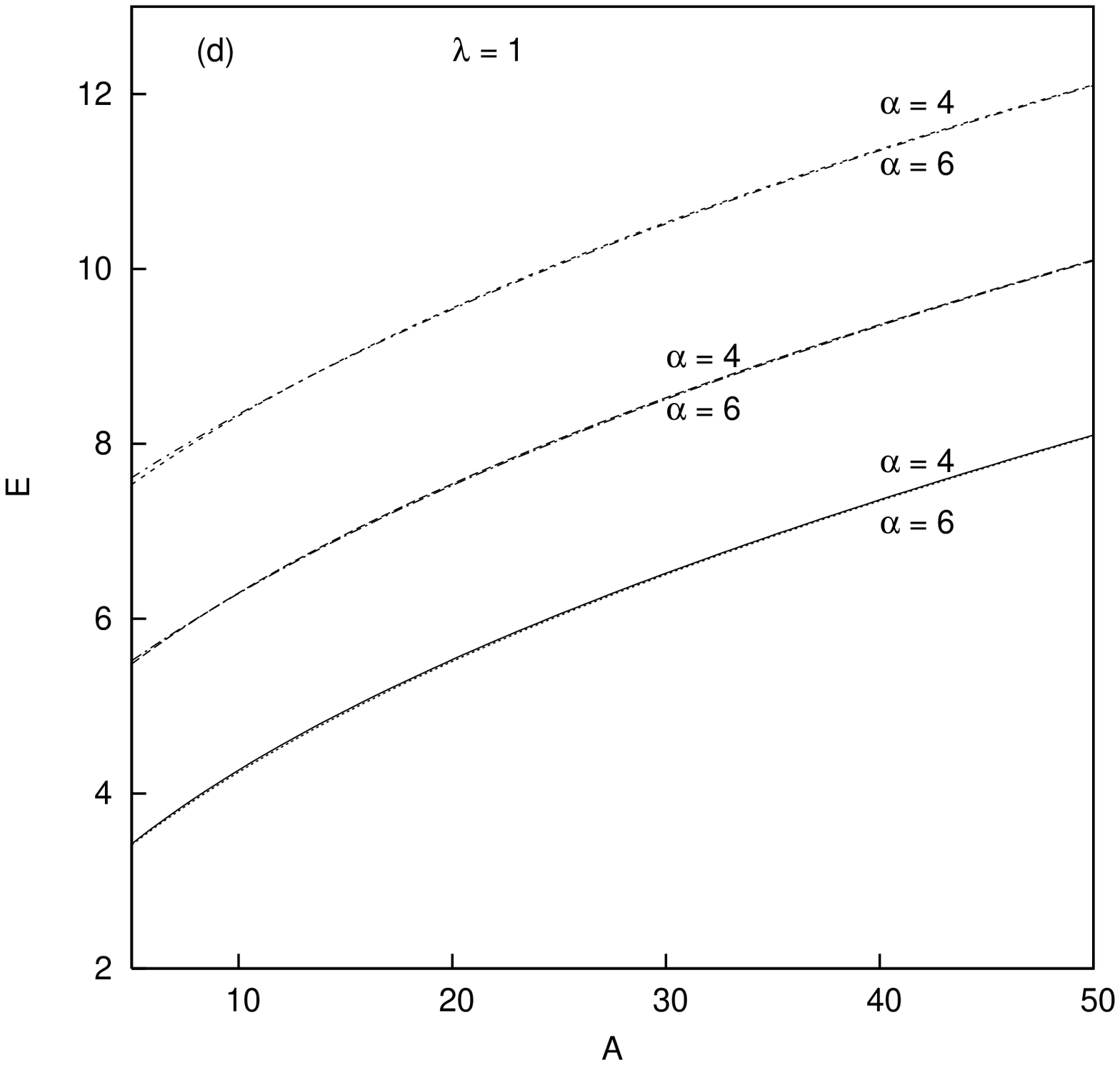}
\end{minipage}
\hspace{0.40in}
\begin{minipage}[b]{0.27\textwidth}\centering
\includegraphics[scale=0.32]{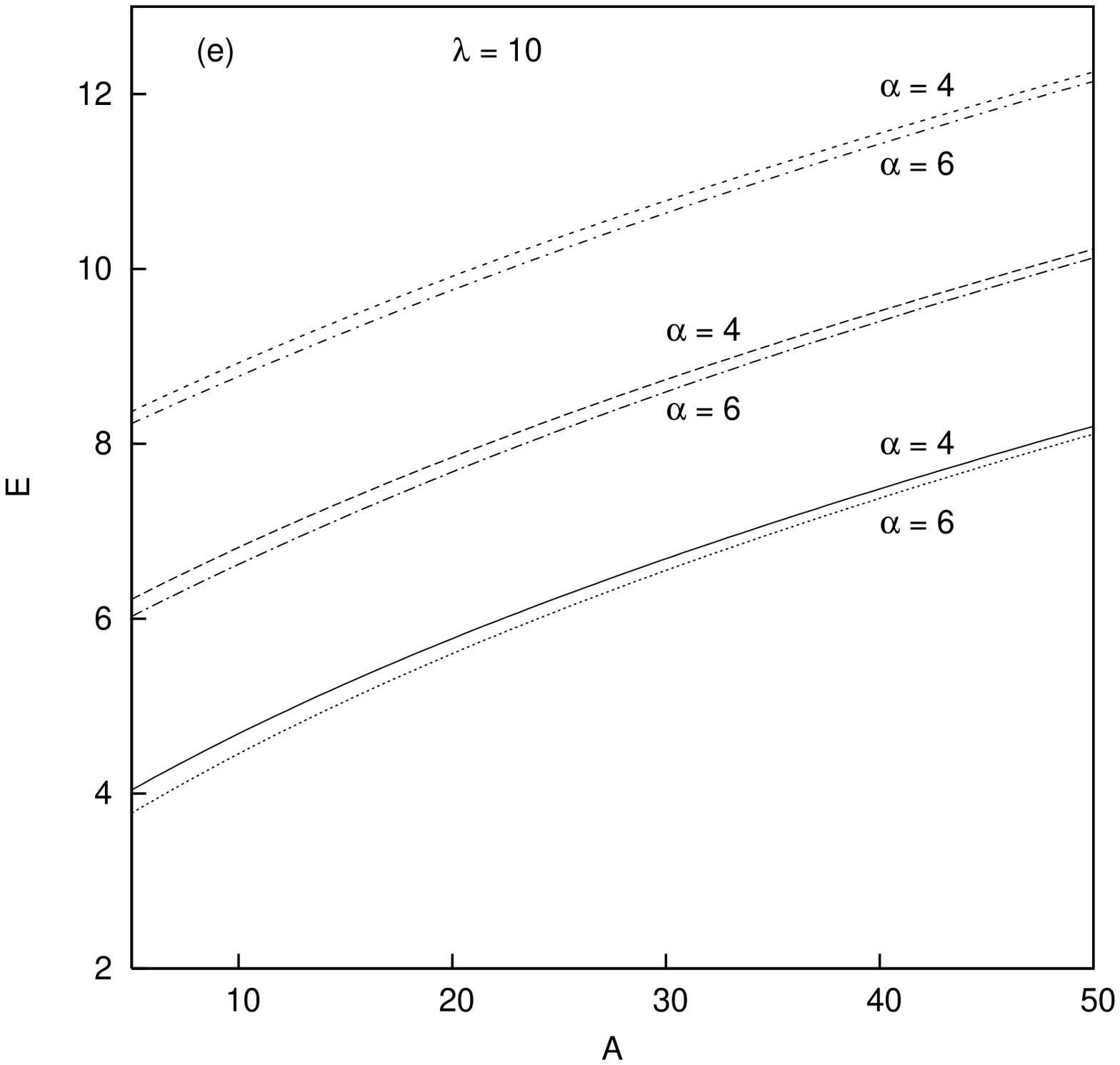}
\end{minipage}
\hspace{0.40in}
\begin{minipage}[b]{0.27\textwidth}\centering
\includegraphics[scale=0.32]{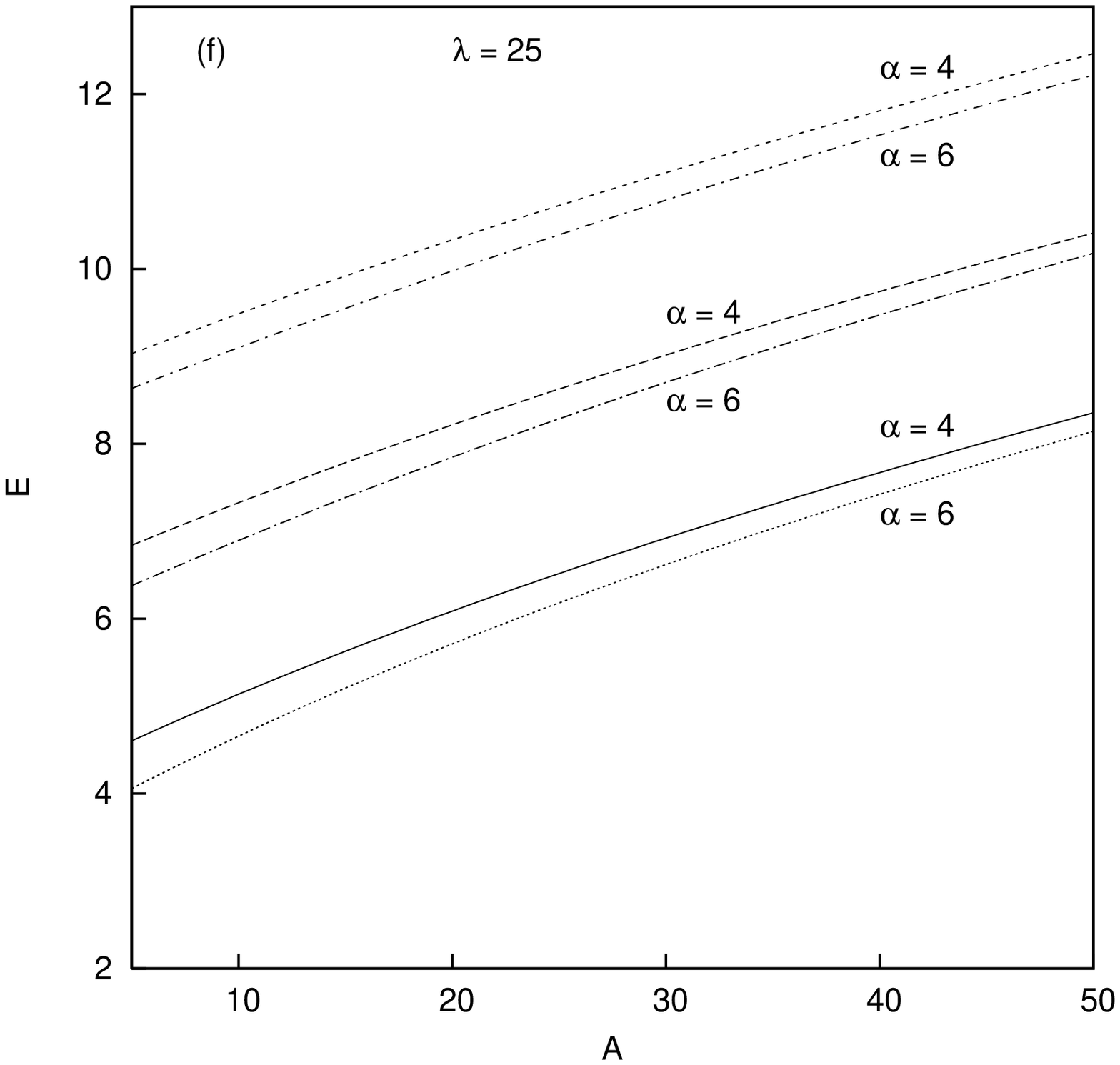}
\end{minipage}
\caption[optional]{Energy variations of a GSHO with respect to $\lambda$ (top) and $A$ (bottom) respectively 
for the first three states belonging to $\ell=0$ having $\alpha=4$ and 6. (a) $A=5$, (b) $A=15$, 
(c) $A=25$; (d) $\lambda=1$, (e) $\lambda=10$, (f) $\lambda=25$. Adopted from \cite{roy08b}.}
\end{figure}

However, to the best of my knowledge, no \emph{direct} results were provided in the above investigations. By
means of GPS method, very accurate ground as well excited states, expectation values, radial densities have 
been reported only lately \cite{roy08b}. A wide range of interaction was considered (both weak and strong 
coupling) for arbitrary values of $n,l$. Energy variations with respect to the two parameters $\lambda, A$ are
displayed at the top and bottom panels in Fig.~1. The first three eigenstates of $l \! = \! 0$ are studied for $\alpha \! = \! 4,6$
respectively with $\lambda$ varied from 0--35. As seen clearly, energy varies rather slowly with respect to $\lambda$ (monotonic 
increase with increase in $\lambda$) for both $\alpha \! = \! 4,6$ (smaller $\alpha$ produces larger effect). In the 
neighborhood of zero $\lambda$, for all values of $A$, both $\alpha$ give very similar energies. As $\lambda$ 
increases, appreciable separation shows up. An increase in $A$ reduces the slope of $\lambda$ vs. E plot, 
eventually becoming flat in (c), signifying even slower variation of E with $\lambda$. The bottom panel 
illustrates $A$ vs E for three $\lambda$ values (1,10,25) in (d)--(f). The same three states of both $\alpha \! = \! 4,6$
make three separate families. For a fixed $\lambda$, energy increases rather promptly relative to the top panel.
Interestingly, $\alpha \! = \! 4,6$ plots seem to be virtually identical for $\lambda \! = \! 1$ for all three states. 
Finally separation between $\alpha \! = \! 4,6$ increases gradually with $\lambda$ as one passes from (d)--(f). For 
further discussion on these results, consult \cite{roy08b}.

\subsection{Screened Coulomb Potentials}
The screened Coulomb potentials, defined as,
\begin{equation}
V(r)=-\frac{Z}{r} \sum_{k=0}^{\infty} V_k(\lambda r)^k
\end{equation}
have found significant importance in many areas in physics, chemistry. This can be used to approximate the
potential experienced by an electron in an atom where other electrons screen the nuclear charge. In the 
form of Debye-H\"uckel potential, it describes the shielding effect in plasmas. An enormous amount of work 
has been done for understanding many fascinating features these potentials exhibit. In the context of atomic 
systems, $Z$ is identified as atomic number while the screening constant $\lambda$ bears different 
significance in different branches. In what follows, we are concerned with the two simple representatives of
the screened Coulomb potential. First one, the Hulth\'en potential \cite{hulthen42,hulthen42a}, given as,
\begin{equation}
V(r)=- \frac{Z\delta e^{-\delta r}}{1-e^{-\delta r}} 
\end{equation}
is one of the most important short-range potentials. It has relevance in nuclear and particle physics 
\cite{durand81,dijk89}, atomic physics \cite{lam71,dutt82}, solid-state physics \cite{lam72,lam78}, chemical 
physics \cite{olson78}, etc. This is also a special case of Eckart potential. Other one, the Yukawa potential 
\cite{yukawa35}, as given below, 
\begin{equation}
V(r)=-\frac{Ze^{-\lambda r}}{r}
\end{equation}
has also found numerous applications in various branches in physics, chemistry, etc.

They show many similarities, e.g., for small $r$, they display Coulomb-like behavior,
\begin{equation}
V(r) \rightarrow -\frac{Z}{r} \ \ \ \ \ \mathrm{for} \ r \rightarrow 0
\end{equation}
whereas they decay monotonically exponentially to zero for large $r$,
\begin{equation}
V(r) \rightarrow O(e^{-\gamma r}) \ \ \ \ \ \mathrm{for} \ r \rightarrow \infty
\end{equation}
Use of a scaling transformation $r \! \rightarrow \! r/Z$ leads to the following well-known relations,
\begin{eqnarray}
E(Z,\gamma) & = & Z^2 E \left( 1,\frac{\gamma}{Z} \right) \nonumber \\
\psi(Z,\gamma,r) &  =  &  Z^{3/2} \psi \left( 1,\frac{\gamma}{Z},Zr \right)   
\end{eqnarray}
Here $\gamma \! = \! \delta \ \mathrm{or} \ \lambda$.
Therefore it suffices to consider only the case of $Z \! = \! 1$, and develop energy and eigenfunctions as a function
of screening parameter $\delta$.
A distinctive feature (in contrast to the Coulomb
case) is that the number of bound states is \emph{limited} because of the presence of screening parameters.  
In other words, bound states exist only for certain values of screening parameter below a threshold limit. 
For Yukawa potential, this value has been estimated quite accurately as 1.19061227$\pm$0.00000004 a.u. \cite{gomes94}. The
former also has an additional property that it offers \emph{exact} analytical solutions for $l \! = \! 0$, not for 
higher partial waves.

\begingroup
\squeezetable
\begin{table} 
\caption {\label{tab:table5}Calculated negative eigenvalues E (a.u.) of various states of Hulth\'en potential 
as function of $\delta$. Numbers in the parentheses denote $\delta_c$ values taken from \cite{varshni90}. 
Asterisks denote exact analytical values. Taken from \cite{roy05}.} 
\begin{ruledtabular}
\begin{center}
\begin{tabular}{llllllll}
State & $\delta$ & \multicolumn{2}{c}{$-$Energy} & State &  $\delta$ &  
\multicolumn{2}{c}{$-$Energy} \\ 
\cline{3-4} \cline{7-8}
   &  & GPS \cite{roy05}   & Literature &  &   & GPS \cite{roy05} & Literature \\   \hline
$1s$(2.0)     & 1.97    & 0.00011249999999 & 0.0001125*  &  
$17s$(0.007)  &  0.005  & 0.0001332288062  & 0.0001332288062*     \\ 
$2p$(0.377)   & 0.35  & 0.00379309814702 & 0.00379309814702\footnotemark[1] &   
$4f$(0.086)   & 0.08  & 0.00135376897143 &                          \\      
$4d$(0.098)   & 0.075 & 0.00383453307692 & 0.00383453307692\footnotemark[1] & 
$6h(0.038)$   & 0.005 & 0.01147020315553                         \\
$8k$(0.021)   & 0.02  & 0.0002027526409  &  &  $10m$(0.013) & 0.01 & 0.0009003110142 \\
\end{tabular}
\end{center}
\vspace{-0.1in}
\end{ruledtabular}
\footnotetext[1] {Ref.~\cite{stubbins93}.}
\end{table}
\endgroup

Great many attempts have been made to calculate the bound-state energy, eigenfunction of these potentials, 
as well their scattering properties.
$l \! \neq \! 0$ states of Hulth\'en potential, correct to any order of $\delta$, have been obtained by using an 
extended version of the analytic perturbation theory \cite{lai80}. Using a non-perturbative approach, an 
extension of Ecker-Weizel approximation was used to obtain analytic closed-form solutions for eigenvalues and
eigenfunctions for arbitrary angular momenta \cite{dutt82}. Latter authors used non-rigorous but intuitive physical
arguments to determine the unknown constants involved in this approximation. To this end, the energy was obtained as, 
\begin{equation}
E_{nl}= -\frac{1}{2}(1/n-n\delta/2)^2 + \frac{1}{8} \delta^2 l(l+1)B [4/n^2 \delta +2 -l(l+1)B/n^2]
\end{equation}
where
\begin{equation}
B=(e-1)^{-1} [(e-1)^2/e -1] = 0.0501438.
\end{equation}
A strong-coupling series was suggested for bound state energies of both these potentials \cite{patil84}
within the WKB approximation, with special emphasis on behavior of energies in the neighborhood of critical region.
Computation of higher-order perturbation theory and summation methods for divergent perturbation series 
was also addressed \cite{popov85}. Later, a path-integral formalism \cite{cai86} 
was put forth for $s$ states, where the exact energy spectrum and normalized $s$-state eigenfunctions were 
obtained from poles of the Green function and their residues. A shifted $1/N$ expansion technique 
\cite{tang87,roy87} was also used for these potentials. Further, an algebraic perturbation method based on Lie algebra 
of the group $SO(2,1)$ which is known to be the dynamical group for a number of spherically symmetric 
potentials, was developed for $s$ states by \cite{dunlap72} and for arbitrary states by \cite{matthys88}. 
A one-parameter variational calculation has also been reported for energies, oscillator strengths \cite{varshni90}. 
These were found to be quite satisfactory in low-screening region, but proved inadequate for higher $\delta$. 
The concept of kinetic potentials was used to construct a global geometrical approximation theory for the
energy spectra of these potentials \cite{hall92}.
A very accurate generalized variational method \cite{stubbins93} was developed for both these potentials. This 
utilized trial functions from a linear combination of independent functions. For Hulth\'en potential, the basis 
functions used assume the following form, 
\begin{equation}
\phi_k=A_kr^ke^{-\beta r}(1-e^{-\delta r})
\end{equation}
where $k \! = \! -1,0,1,2,\cdots $ for $s$ states and $k \! = \! 0,1,2,\cdots $ for $l \! \ne \! 0$ states. $\beta$ is a 
variational parameter determined by minimizing the energy for a given state and basis size. The constant
$A_k\!=\!(2\beta+\delta)^{k+1}/\sqrt{(2k+2)!}$ is included to prevent numerical overflow. For Yukawa
potential, the basis takes the following form,
\begin{equation}
\psi_k\!=\!B_k r^k e^{-\beta r/2}, \ \ \ \ k=0,1,2,\cdots 
\end{equation}
where $B_k, \beta$ are normalization constant and variational parameter respectively.
An improved variational scheme was proposed \cite{fessatidis98}, wherein a set of variational parameters 
was introduced into the trial wave function to form a family of independent functions. This potential has also been 
dealt using supersymmetric quantum mechanics and first-order perturbation theory \cite{gonul00}. Based on
the small and large $r$ behavior, threshold and asymptotic properties of energy and eigenfunctions of these
potentials were examined critically in \cite{patil01}.

\begin{figure}
\begin{minipage}[c]{0.40\textwidth}
\centering
\includegraphics[scale=0.45]{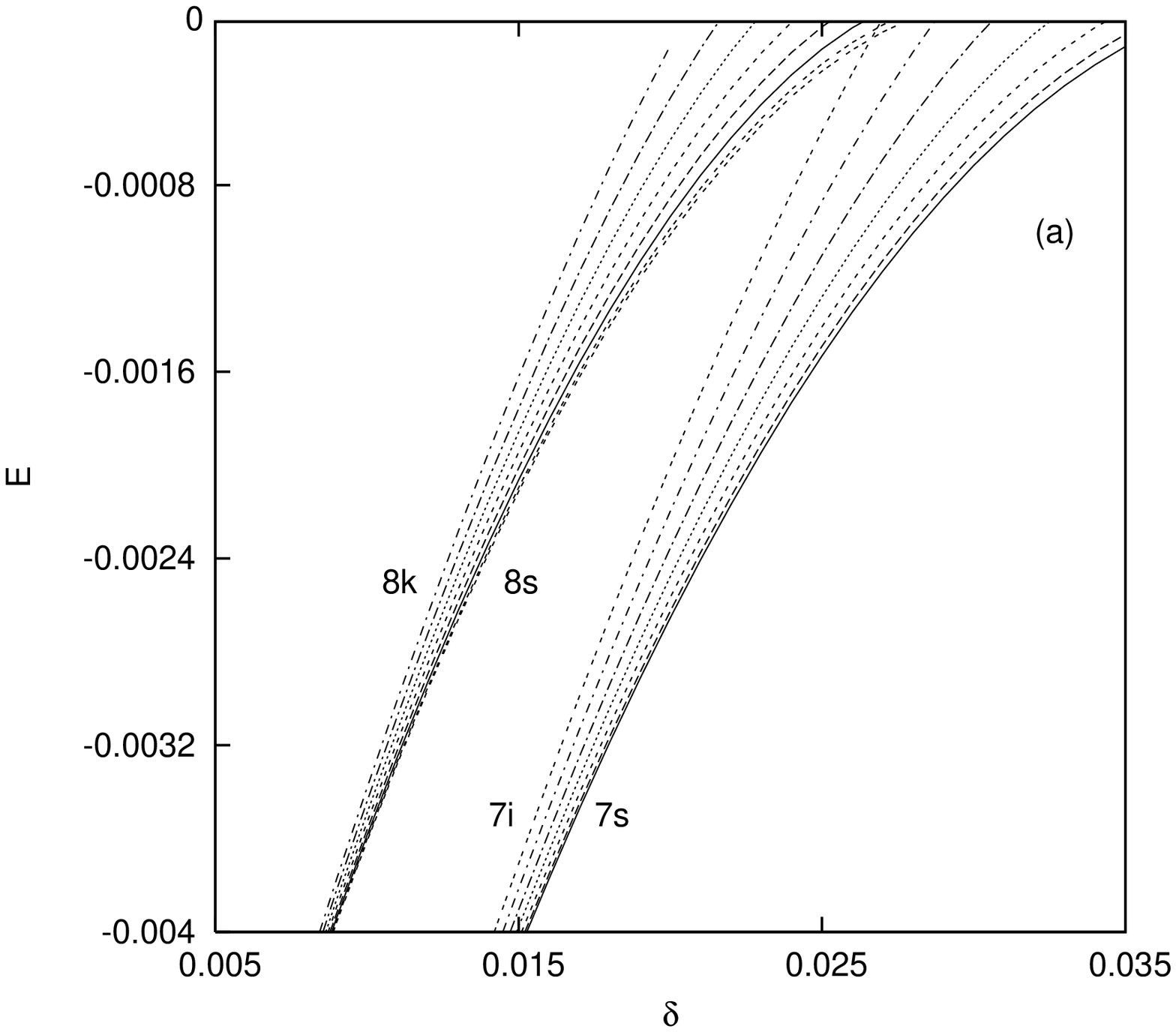}
\end{minipage}%
\hspace{0.5in}
\begin{minipage}[c]{0.40\textwidth}
\centering
\includegraphics[scale=0.45]{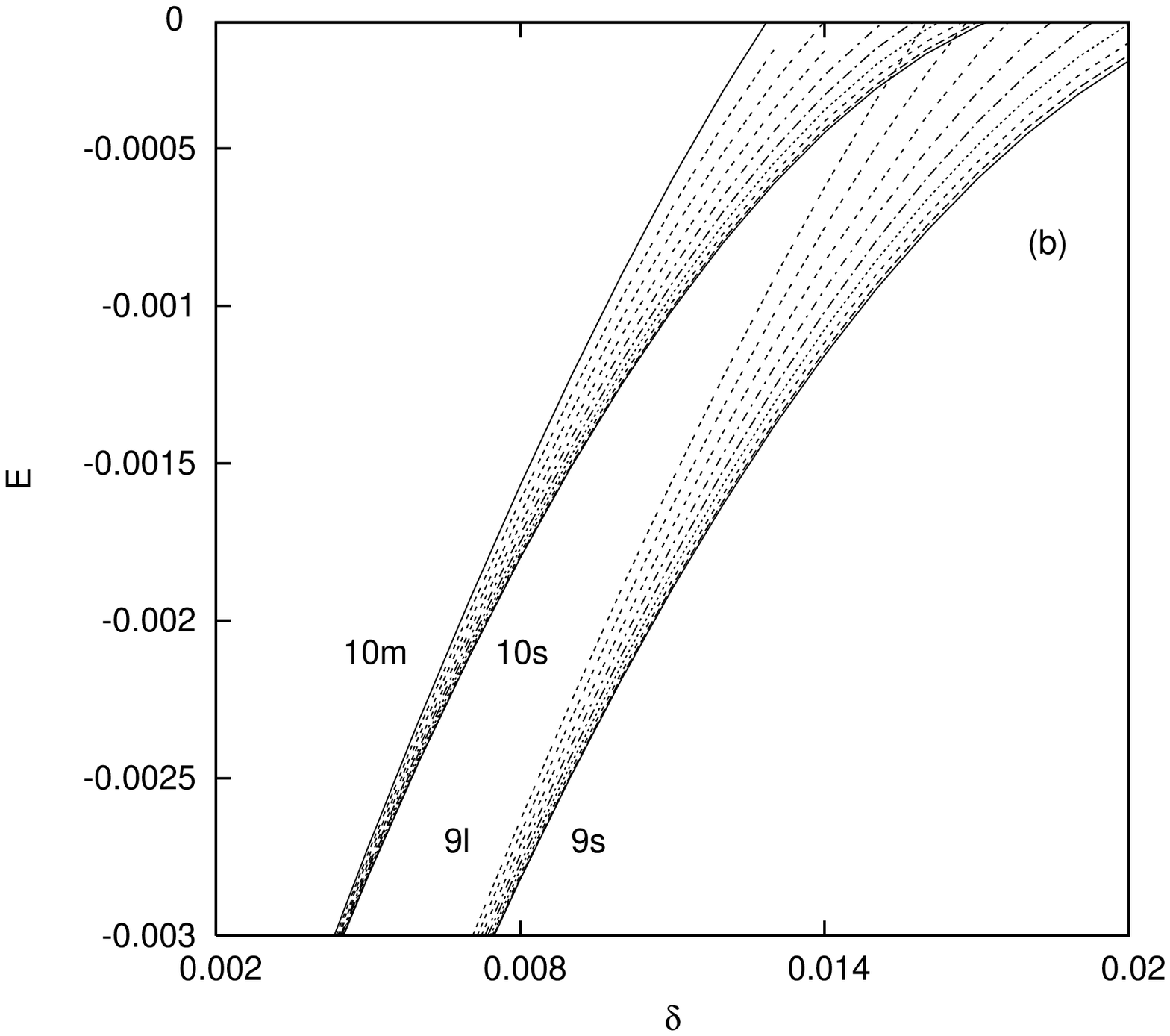}
\end{minipage}%
\caption{Energy eigenvalues (a.u.) of the Hulth\'en potential for (a) $n=7,8$ and 
(b) $n=9,10 $ levels respectively as a function of $\delta$ in the vicinity of 
zero energy. Taken from \cite{roy05}.}
\end{figure}

First-order perturbative calculations using Hulth\'en potential as unperturbed potential, was employed for
$s$ states of Yukawa potential \cite{lam71}. Later, motivated by this, using solutions to a Hulth\'en-like 
effective potential as variational trial functions, nonzero $l$ states of Yukawa potential were obtained 
\cite{greene76} with reasonable accuracy. It was soon realized that these trial functions provide better 
variational energies, wave functions with fewer parameters than frequently used hydrogenic or Slater-type 
functions. Detailed variational results were presented for lowest 45 eigenstates \cite{roussel74}; also 
probabilities for spontaneous emission in dipole approximation was studied as a function of screening 
length, for transitions between six lowest states. A combined Pad\'e approximant ([6,6], [6,7]) to the 
perturbation series for Yukawa potential was presented \cite{lai81}. A shifted $1/N$ expansion 
\cite{imbo84,moreno84,chatterjee85} as well as an improved expansion was put forth \cite{christiansen89}. 
These offered convergence, better 
than usual $1/N$ expansion. Through a combined perturbation theory and continued fractions-Pad\'e
approximations at large order \cite{vrscay86}, very accurate bound states were obtained. A linear combination 
of atomic orbitals (LCAO) calculation was performed for ground states as function of $\lambda$ within a 
variational framework (using Slater basis functions) \cite{gomes94}.

Accurate numerical methods have been developed for these potentials. All the 45 eigenstates $1s$ through 
$n \! = \! 9,l \! = \! 8$ were numerically investigated for wider $\lambda$ for Yukawa potential nearly four decades ago
with reasonably good accuracy \cite{rogers70}. Another of them \cite{nunez93} consists in 
solving the Dirichlet problem in a box with radius $n$ by a Ritz method. The convergence to the
eigenfunctions in the norm of Hilbert space $L_2(0,n)$ was proved to be guaranteed.

\begingroup
\squeezetable
\begin{table}  
\caption {\label{tab:table6}Comparison of the negative eigenvalues (a. u.) of Yukawa potential as function 
of $\lambda$. Numbers in the parentheses denote $\lambda_c$ values. See \cite{roy05} for details.} 
\begin{ruledtabular}
\begin{center}
\begin{tabular}{lcllccll}
State & $\lambda$ & \multicolumn{2}{c}{$-$Energy} & State &  $\lambda$ & 
\multicolumn{2}{c}{$-$Energy}   \\ 
\cline{3-4} \cline{7-8}
   &  & GPS \cite{roy05} & Literature  &  &   & GPS \cite{roy05} & Literature \\    \hline 
$4d$(0.0581)& 0.01 & 0.02222779248980 & 0.02222779248980\footnotemark[1]  & 
$4d$        & 0.055 & 0.00049188376726 &                      \\
$9s$(0.016)  & 0.01  & 0.0005858247612 & 0.000585\footnotemark[2] & $10s$ & 0.005 & 0.0015083559307 &   \\
$9p$(0.015)  & 0.01  & 0.0005665076261 & 0.000565\footnotemark[2] & $10p$ & 0.005 & 0.0015009235029 &   \\
$9l$(0.0094) & 0.005 & 0.0021291265596 & 0.00213\footnotemark[2]  & $10l$ & 0.005 & 0.0012296811835 &   \\
$10m$        & 0.005 & 0.0011557947569 &          & $17s$ & 0.001 & 0.000919120394  &   \\
\end{tabular}
\begin{tabbing}
$^{\mathrm{a}}\!$ Ref.~\cite{stubbins93}. \hspace{0.3in} \= $^{\mathrm{b}} \!$ Ref.~\cite{rogers70}. 
\end{tabbing}
\end{center}
\vspace{-0.2in}
\end{ruledtabular}
\end{table}
\endgroup

It may be mentioned that although there are decent number of high-quality results available for these 
interactions in the literature in \emph{weak}-coupling regions as well as \emph{lower} states, there is 
a scarcity of such results for \emph{stronger} coupling and for \emph{higher} states. In a GPS study 
\cite{roy05}, these potentials were calculated very accurately for arbitrary field strengths with special 
emphasis on these issues. Sample Hulth\'en potential results from GPS method are collected in 
Tables V. Numbers in the parentheses denote respective critical screening constants. Literature values 
are quoted for comparison, wherever possible. As discussed in \cite{roy05}, either for higher states or high screening 
constants, larger $R$ is needed, whereas eigenvalues are apparently less sensitive with respect to the number
of grid points $N$. In Table V, the $s$ states are presented in the strong $\delta$ region. Exact analytical
results (denoted by asterisks), available for $l=0$ states of Hulth\'en potential are given as, 
\begin{equation}
E_n^{exact}=-\frac{\delta^2}{8n^2} \left[ \frac{2}{\delta}-n^2 \right]
\end{equation}
with $n^2 < 2/\delta$. Our calculated values for $s$ states completely coincide with exact analytical results 
for all states (up to $n \! = \! 17$) covering a whole range of interaction. The $\delta_c$ for $l \! = \! 0$ state is given 
exactly by the simple relation $\delta_c \! = \! 2/n^2$, whereas for $l \! \ne \! 0$ this is approximated by the 
analytical expression 
\cite{patil84},
\begin{equation}
\delta_c=1/[n\sqrt{2} + 0.1645l +0.0983l/n]^2
\end{equation}
which offers good agreement with numerically determined values \cite{varshni90}. Excellent agreement with literature
results has been observed for all the states. As seen, a uniform accuracy is 
achieved for all states encompassing a large range of interaction, unlike some previous calculations which 
encountered difficulties in the strong-coupling region. For weaker coupling, present results are superior to 
all other results except the variational work of \cite{stubbins93}. However, in the stronger region, GPS results 
are superior. We have enlarged the coupling region from all other previous works and current results are so far 
the most accurate ones in the neighborhood of critical $\delta$. Additionally, Figure 2 depicts the variation of energies with
respect to $\delta$ for all states belonging to $n \! = \! 7,8$ (a) and $n \! = \! 9,10$ (b) in the neighborhood of 
zero energy. For small $n$, there is good 
resemblance of energy orderings with those of Coulomb potentials; however, this scenario changes with an increase 
in $n$. In that case, significant deviation is noticed as well as complex level crossing observed in the vicinity of 
zero energy, which makes their accurate determination quite difficult. This is more dramatic for latter (e.g., 
$9k$, $9l$ mixing heavily with $10s$, $10p$, $10d$, $10f$ at around $\delta =0.015-0.017$). Besides, for a given 
$n$, separation between states with different $l$ increases with $\delta$. Further discussion on results including radial
densities, expectation values etc., could be found in \cite{roy05}.

\begin{figure}
\begin{minipage}[c]{0.40\textwidth}
\centering
\includegraphics[scale=0.45]{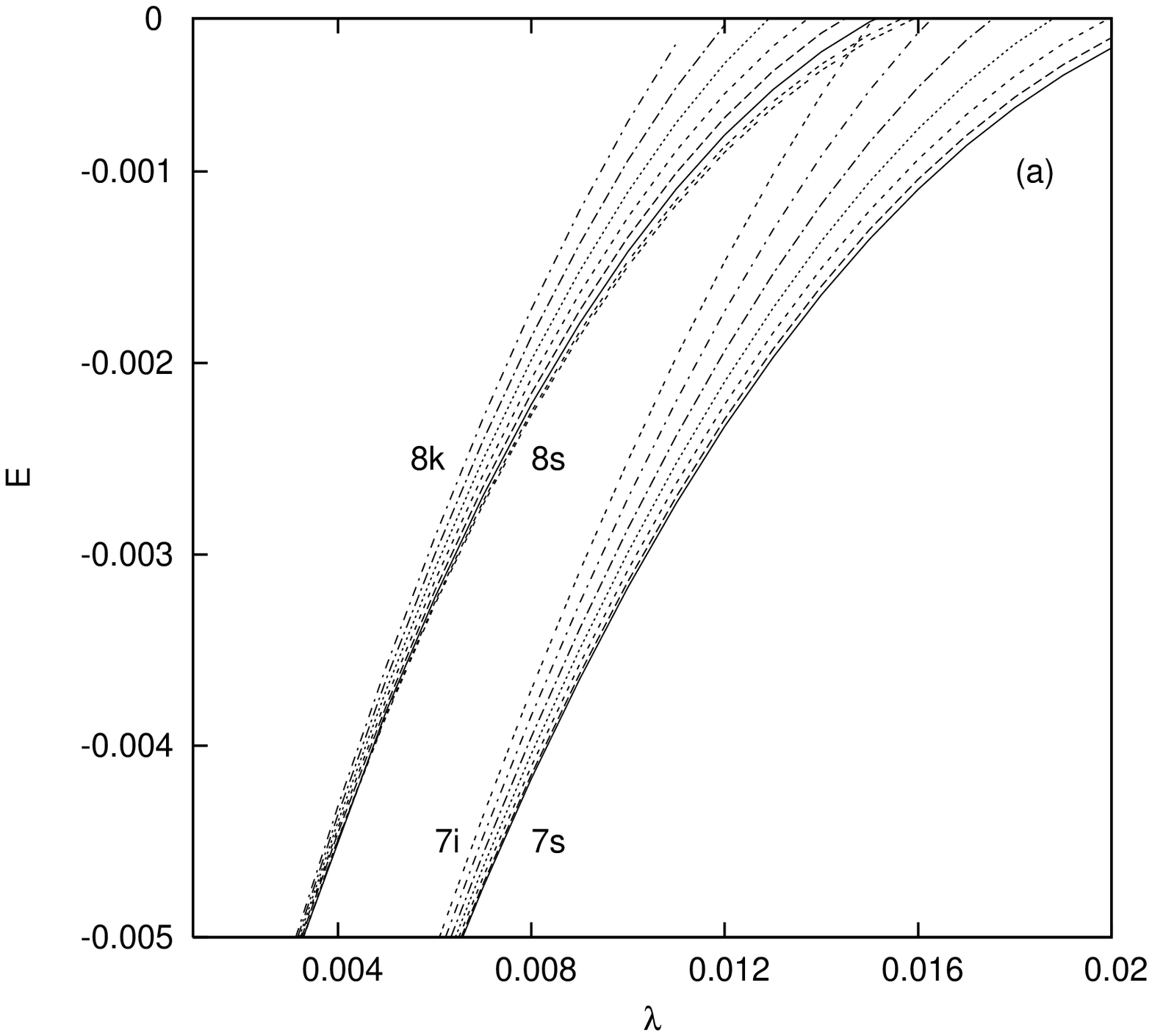}
\end{minipage}%
\hspace{0.5in}
\begin{minipage}[c]{0.40\textwidth}
\centering
\includegraphics[scale=0.45]{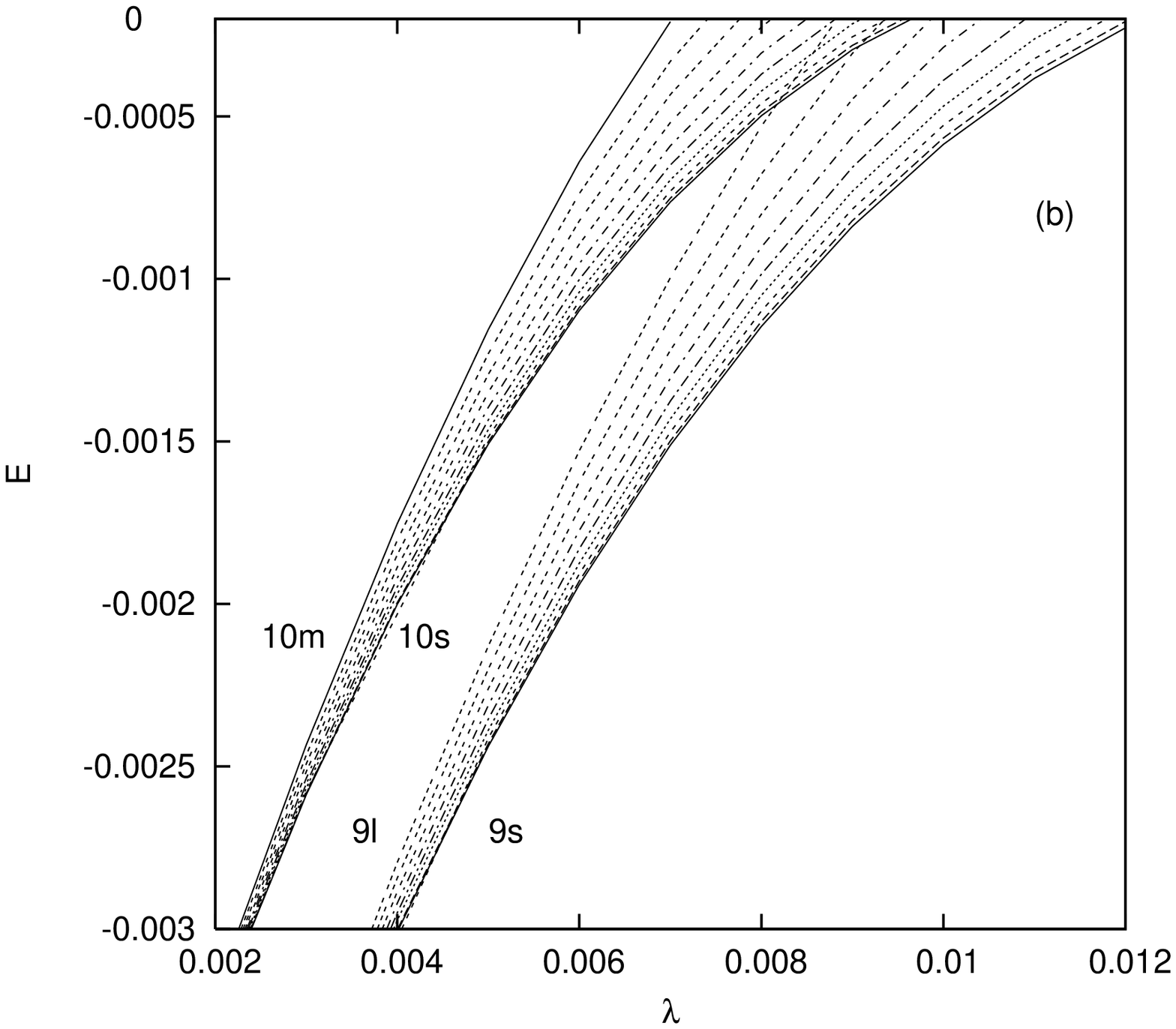}
\end{minipage}%
\caption{Energy eigenvalues (a.u.) of the Yukawa potential for (a) $n \! = \! 7,8$ and (b) 
$n \! = \! 9,10$ (right) levels respectively as a function of $\lambda$ in the vicinity of
zero energy. Adopted from \cite{roy05}.}
\end{figure}
Let us now turn our focus on the Yukawa potential. Similar results as those Hulth\'en potential, are presented
in Table VI and Fig.~3. No analytical expressions are available for critical screening constants in this case; 
numerically determined $\lambda_c$s, available from \cite{rogers70}, are quoted, wherever possible. As in the 
previous case, here also very good agreement is observed for all states with the best theoretical results. Once 
again, a large range of interaction (both weak, strong) as well as very high states are considered. As in Hulth\'en 
potential, results of \cite{stubbins93} are more accurate than ours in the small $\lambda$ region, but for 
stronger couplings, present results are superior. Some of the higher states have not been calculated by any method
other than those of \cite{rogers70,patil01}; our results significantly improve those. Considering all these, 
present results appear to be the most accurate and reliable for all the states (except $1s$, $2s$) in regions
close to $\delta_c$. As $n,l$ increase, accurate calculation of these states become progressively difficult; 
for $7s-9l$ states only two attempts \cite{rogers70,imbo84} are known other than the present one. In Fig.~3, 
dependence of energy orderings on $\lambda$ is shown vividly in the neighborhood of $E \! = \! 0$ for $n \! = \! 7,8$ (a) and 
$n \! = \! 9,10$ (b). Very similar conclusions as in Fig.~2 can be drawn: (i) as $n$ increases energy ordering tends 
to be more complex with the possibility of level crossing becoming higher (ii) energy splitting between states 
with different $l$ increases with an increase in $\lambda$ for a particular $n$. No attempts are known for any
of these states with $n >9$ other than the present one and these may constitute a useful reference for future
studies. As an illustration, also some very high-lying states ($17s$) are included in this table. Finally, we 
note that eigenvalues of Coulomb, Hulth\'en and Yukawa potentials are known to satisfy the following relation,
\begin{equation}
E_n^{\mathrm{coulomb}} \leq E_{n,\ell}^{\mathrm{Hulthen}}(\delta)
\leq E_{n,\ell}^{\mathrm{Yukawa}}(\lambda)
\end{equation}
This has been found to be true for all the states considered here. For many other interesting features on these
potentials, see \cite{roy05}.

\subsection{Power-law and Logarithmic potentials}
The power-law and logarithmic potentials have found wide-spread applications in particle physics 
\cite{quigg77,jena86,akcay96,jena01}. The orbital structure of logarithmic potential has been 
studied in the context of self-consistent modeling of triaxial systems (such as elliptical galaxies), 
bars in the centers of galaxy discs \cite{schwarszchild79}, global dynamics \cite{stoica03}, etc. The 
Coulomb plus power-law potential also serves as a non-relativistic model for the principal part of a 
quark-quark interaction \cite{ciftci03}. The potential is given by $V(r) \! = \! A \ \mathrm{sgn} (\nu)r^{\nu}$, 
where $r \! = \! ||r||$, $A \! > \! 0$ and $\nu \! \ne \! 0$. For dimension $N \! = \! 1$, $\nu \! > \! -1$ and 
for higher dimensions 
$N \! \ge \! 2$, $\nu \! > \! -2$. For $\nu \! = \! 0$, we have $V(r) \! = \! A \ln (r)$, with $A \! > \! 0$. It is possible 
to include 
logarithmic potential as a limiting case of the power potentials if in place of the potential family 
$f(r) \! = \! \mathrm{sgn}(\nu)r^{\nu}$, we use $V(r,\nu) \! = \! (r^{\nu}-1)/{\nu}$, whose limit as $\nu \! \rightarrow \! 0$ 
is $V(r,0) \! = \! \ln (r)$. Eigenvalues of the power-law potential $E_{nl}^N$ can be labeled by two quantum 
numbers; the total angular momentum $l \! = \! 0,1,2,\cdots ,$ and a `radial' quantum number $n \! = \! 1,2,3,\cdots $, 
which represents 1 plus number of nodes in the radial part of wave function. The eigenvalues in 
$N \ge 2$ spatial dimensions has degeneracy 1 for $l=0$ and for $l >0$, the same is given by,
\begin{equation}
\Lambda (N,l)=\frac{(2l+N-2)(l+N-3)!}{l!(N-2)!}, \ \ \ N \ge 2, l>0
\end{equation}
The well-known hydrogenic atom and harmonic oscillator constitute the two exactly solvable cases in $N$ 
dimensions corresponding to $\nu=-1,2$ respectively,
\begin{eqnarray}
E_{nl}^N(-1) & = & -[2(n+l+N/2-3/2)]^{-2}, \ \ \ N \ge 2  \nonumber \\
E_{nl}^N(2) & = & 4n+2l+N-4, \ \ \ \    \ \ \ N \ge 2 
\end{eqnarray}
Also analytical solutions for linear potential in 1D, as well as $s$ states in 3D could be found from the
zeros of Airy function \cite{quigg79}.

Numerous attempts \cite{biswas73,nieto79a,quigg79,crandall82,sukhatme83,imbo84a,maluendes86,hall89,hall97,
hall03,ciftci03a} have been made in the past years to 
examine many interesting properties of their solutions utilizing an array of methodologies. For example, ground and 
excited energy levels of the generalized 1D anharmonic oscillator characterized by the potential 
$V(x) \! = \! x^2+\lambda x^{2m}, \ m \! = \! 2,3$ were calculated non-perturbatively \cite{biswas73} by a Hill determinant 
method. A WKB approximation \cite{quigg79} was proposed. It was proved \cite{nieto79a} that the eigenvalues 
$E_n \! = \! E_{n0}^1$ of power-law potentials in 1D 
increase with $n$ at a higher rate for a greater $\nu$. However for any $\nu$, this increase never attains 
$n^2$. In general, the dependence of $E_{nl}^N$ on the coupling parameter $A$ may be established by elementary 
scaling arguments by replacing $r$ by $\sigma r$. Then one finds that, 
\begin{equation}
E_{nl}^N (A) =A^{2/(\nu+2)} E_{nl}^N(1)
\end{equation}
Thus without any loss of any generality, one can limit further discussion on the case of unit coupling 
$A \! = \! 1$. Lower and upper analytic bounds for ground states were developed for power law potentials 
\cite{crandall82}. The shifted $1/N$ expansion technique, with some of its variants (such as a modified or
large-order expansion) \cite{sukhatme83,imbo84a,maluendes86} has been quite successful. Dependence of 
eigenvalues on the power parameter $\nu$ has been studied by spectral geometrical arguments \cite{hall89}. 
Bounds of eigenvalues for polynomial potentials in $N$ dimension have also been studied semi-classically 
\cite{hall97,hall03}. A variational method \cite{ciftci03a} was developed for eigenvalues and eigenfunctions;
it provided good results for small $\nu$, but suffered in the higher $\nu$ regions. 

GPS method \cite{roy04a} has produced very accurate results for both these potentials for states with arbitrary
quantum numbers $n,l$. A detailed comparison with literature results has been made in \cite{roy04a}, from which 
it is abundantly clear that the 
present scheme offers results which are considerably better than the existing results available. Moreover, as 
in the previous occasions, here also we obtain both low as well as higher states with equal ease and accuracy. 
As an illustration, Table VII compares our calculated eigenvalues for two potentials $V(r) \! = \! -2^{1.7}r^{-0.2}$ and
$V(r) \! = \! 2^{7/2}r$, which have been examined earlier quite extensively by other workers. Selected states with 
$n \leq 4$ and $l \leq 3$ are reported here. In the former case, accurate results are not available 
for comparison and the present results match closely to all of the results available, while for for the latter case, 
GPS results are in excellent agreement with the accurate values of \cite{maluendes86}. For a more complete discussion,
see \cite{roy04a}.  

\begingroup
\squeezetable
\begin{table}  
\caption {\label{tab:table7}Comparison of eigenvalues of the two power-law potentials $V(r) \! = \! -2^{1.7}r^{-0.2}$ 
(left) and $V(r) \! = \! 2^{7/2}r$ (right) for several $n,l$ values. See \cite{roy04a} for details.} 
\begin{ruledtabular}
\begin{center}
\begin{tabular}{llllll}
 \multicolumn{2}{c}{Energy}& $n$ & $\ell$ & \multicolumn{2}{c}
{Energy} \\ 
\cline{1-2} \cline{5-6}
 GPS~\cite{roy04a}  & Literature  &  &  &  GPS~\cite{roy04a} & Literature \\
\hline
$-2.68602822$ & $-2.68601$\footnotemark[1], $-2.6859$\footnotemark[2],
$-2.686$\footnotemark[3] & 0 & 0 & 9.352429641 &
9.352429643\footnotemark[4], 9.3524296418\footnotemark[5] \\
$-2.04431800$ & $-2.04658$\footnotemark[1], $-2.0440$\footnotemark[2],
$-2.044$\footnotemark[3] & 2 & 0 & 22.08223931 & 
22.08223931\footnotemark[4], 22.0822393124\footnotemark[5] \\
$-1.81414352$ &     & 4 & 0 & 31.77653434 &                \\
$-2.02906490$ & $-2.02906$\footnotemark[1], $-2.0291$\footnotemark[2],
$-2.029$\footnotemark[3] & 0 & 3 & 20.20370253 &           \\
$-1.90486674$ & $-1.90491$\footnotemark[1], $-1.9049$\footnotemark[2],
$-1.905$\footnotemark[3] & 1 & 3 & 25.32846149 &           \\
$-1.73987512$ &          & 3 & 3 & 34.38846804 &           \\
\end{tabular}
\begin{tabbing}
$^{\mathrm{a}}\!$ Ref.~\cite{imbo84a}. \hspace{0.2in} \= $^{\mathrm{b}} \!$ Ref.~\cite{ciftci03a}. 
\hspace{0.2in} \= $^{\mathrm{c}} \!$ Numerical results, from \cite{imbo84a}. \hspace{0.2in}  \= 
$^{\mathrm{d}} \!$ Ref.~\cite{maluendes86}. \hspace{0.2in} \= $^{\mathrm{e}} \!$ Exact values, as quoted in \cite{maluendes86}. 
\end{tabbing} 
\end{center}
\vspace{-0.2in}
\end{ruledtabular}
\end{table}
\endgroup

\subsection{Rational Potentials}
Rational, also termed as non-polynomial oscillator (NPO) potentials, defined by, 
\begin{equation}
V(r)=r^2+\frac{\lambda r^2}{1+gr^2}; \ \ \ g>0, \ \lambda \in (-\infty, \infty)
\end{equation}
has been one of the most important model systems in quantum mechanics. The relevance in nonlinear Lagrangian
field theory \cite{risken67}, laser theory (as the reduction of Fokker-Planck equation of a single-mode laser
under suitable conditions) and nonlinear optics \cite{haken70} was pointed out long times ago. The Schr\"odinger 
equation with such an interaction Lagrangian is analogous to a zero-dimensional field theory with a nonlinear 
Lagrangian in elementary particle physics \cite{biswas73,biswas78}. Also the 3D analogue was found to produce a
sequence of energy levels which is identical to that occurring in the shell model of nucleus \cite{varshni87}.
It may be noted that for either of the following situations, $\lambda \! = \! 0$ or $\lambda \! = \! g \! = \! 0$ or 
$g \! << \! \lambda$ or large $g$, the solution behaves as the harmonic oscillator. 

The potential in 1D has generated considerable interest among the theoreticians as evidenced by numerous works 
employing a wide range of methodologies such as variational method, perturbation theory, semi-numerical as well 
as purely numerical methods. It is possible to obtain \emph{exact} eigenvalues and eigenfunctions of ground and
higher states provided the potential parameters obey certain specific relations between them. In \cite{whitehead82}, 
existence of a class of solutions (in terms of terminating polynomials or Sturmians of Schr\"odinger equation with
potential $x^2-\lambda/\{g(1+gx^2)\}, -\infty <x < \infty$) was found out when certain algebraic relations between 
$g,\lambda$ are satisfied. In another attempt \cite{flessas82}, eigenfunctions were expressed as definite integrals
whereas eigenvalues by means of a limiting procedure. Exact even- and odd-parity solutions in the form of products
of exponentials and polynomial of x have been investigated \cite{chaudhuri83}. For small $\lambda' (=\lambda/g)$, 
the eigenvalues were given by,
\begin{equation}
E_n= 2n+1+\frac{1}{2}\lambda' -\lambda'(\sqrt{\pi} \ 2^n n!)^{-1} I_n
\end{equation}  
where 
\begin{equation}
I_n= \int_0^{\infty} \exp (-x^2) H_n^2(x) (1-gx^2)/(1+gx^2) dx
\end{equation}
with $n=0,1,2,\cdots$, and $H_n(x)$ is a Hermite polynomial of order $n$. Similarly, the first four eigenvalues 
(for large $g$) were given as, 
\begin{eqnarray}
E_0 & = & 1+ \lambda' \left( 1-\sqrt {\pi} \ g^{-1/2} +\frac{5}{2}g^{-1} \right)  \nonumber \\
E_1 & = & 3+ \lambda' \left( 1-\frac{3}{2} g^{-1} +2 \sqrt{\pi} \ g^{-3/2} \right)  \nonumber \\
E_2 & = & 5+ \lambda' \left( 1-\frac{1}{2} \sqrt {\pi} \ g^{-1/2} +\frac{9}{4}g^{-1} \right)  \nonumber \\
E_3 & = & 7+ \lambda' \left( 1-\frac{3}{2} g^{-1} +\frac{3}{2} \sqrt{\pi} \ g^{-3/2} \right) 
\end{eqnarray}
It is possible to supersymmetrize the non-polynomial interaction and in that case, one may find as many as 
exact analytical solutions (corresponding to ground states of different supersymmetric quantum mechanical system)
one wishes \cite{roy87a}. Existence of conditionally exact solutions for 1D NPO has been studied by other authors 
as well \cite{flessas81,varma81,roy88}. Possibility of an infinite set of exact solutions of both odd- and even-parity, 
which could be expressed in terms of a product of exponential and polynomial functions of $x^2$ for specific 
relations between $\lambda,g$, has been explored in \cite{blecher87,gallas88,berghe89} as well.  

\begingroup
\squeezetable
\begin{table} 
\caption {\label{tab:table8}Comparison of some lowest ($n_r=0$) eigenvalues E (times 2 in a.u.) of 
3D NPO for several $g$ and $\lambda$ with literature data for $\ell=0,1,2$. Exact analytical values 
are referred by asterisks \cite{saad06,roy88}. GPS results are quoted from \cite{roy08a}.} 
\begin{ruledtabular}
\begin{center}
\begin{tabular}{crrrrrrrrr}
        &  &     & \multicolumn{2}{c}{ E}  &  &  & & \multicolumn{2}{c}{ E}  \\
\cline{4-5}  \cline{9-10}
$\ell$ & $g$ & $\lambda$ & GPS  & Ref.   &  $\ell$ & $g$  & $\lambda$  & GPS & Ref.      
\\  \hline
 0   & 0.1  & $-$0.46   & 2.4000000000000     & 2.400000000000\footnotemark[1],2.4*   &
 1   & 0.1  & $-$0.5    & 3.9999999999999     & 4.000116\footnotemark[2],4*          \\
 0   & 1    & $-$10     & $-$3.000000000000   & $-$3.000000000000\footnotemark[1],$-$3*  &
 1   & 0.01 & $-$0.041  & 4.9000000000000     & 4.899974\footnotemark[2],4.9*        \\
 2   & 1    & $-$18     & $-$6.999999999999   & $-$7.000000000000\footnotemark[1],$-$7*   & 
 2   & 10   & $-$1440   & $-$133.000000000000 & $-$133.000000000000\footnotemark[1],    \\
     &      &           &                     &                                           &
     &      &           &                     & $-$133*       \\
\end{tabular}
\begin{tabbing}
$^{\mathrm{a}}\!$ Variational calculation \cite{saad06}. \hspace{0.3in} \= $^{\mathrm{b}} \!$ 
Shifted 1/N expansion result \cite{roy88a}. 
\end{tabbing}
\end{center}
\vspace{-0.2in}
\end{ruledtabular}
\end{table}
\endgroup

The literature for 1D NPO is vast; only some of the most important ones are cited here chronologically. A 
non-perturbative method, in conjunction with a Ritz variational (with a Hermite polynomial basis) method was 
developed for ground and first two excited states with reasonable success \cite{mitra78}. Using perturbation
theory, asymptotic expansions were derived \cite{kaushal79} for energies and eigenfunctions in the range of 
small $g$ and large $\lambda$. A combined variational
and perturbation theory with properly scaled harmonic oscillator functions as basis set was used as well 
\cite{bessis80}. A Hill-determinant method \cite{hautot81} was proposed. Ground and first three excited states
of the interaction potential were obtained by forming a [6,6] Pad\'e approximant to the energy perturbation 
series through a hypervirial relation \cite{lai82}. A perturbed ladder operator method \cite{bessis83} was applied 
to the resolution of perturbed harmonic oscillator wave equation for cases when the perturbation is 
expandable in a convergent series of Hermite polynomials. Some other methods are: (i) variety of finite-difference 
approaches of different flavor \cite{fack85,witwit92,witwit96} with differing accuracy (ii) quasi-polynomial
solutions with the help of first Heun confluent equation or spheroidal Heun equation \cite{marcilhacy85} (iii) 
an algebraic perturbation theory, based upon 
the SO(2,1) dynamical group and a tilting transformation, found to be quite successful for eigenfunctions,
eigenvalues in the small $g$ region \cite{fack86} (iv) supersymmetric as well ordinary WKB method \cite{roy88}
(v) a mixed-continued fraction algorithm \cite{scherrer88} (vi) an analytic continuation procedure \cite{hodgson88} 
using a Taylor series, which produces very accurate energies and wave functions (vii) perturbation theory with 
mixed hypervirial and Hellmann-Feynmann theory \cite{witwit92a} (viii) a composite of modified Hill-determinant 
as used in \cite{chaudhuri83}, incorporating an operator method and a vector recurrence \cite{agrawal93}, gives
quite accurate solutions (ix) variational bounds via Rayleigh-Ritz theorem \cite{stubbins95} (x) a quadrature 
discretization technique \cite{chen98} (xi) purely numerical approach \cite{ishikawa02}, etc. While most of these
focus on 1D case, some (such as \cite{scherrer88}) deal with both 1D plus 3D and/or N dimensions. 

\begingroup
\squeezetable
\begin{table}
\caption {\label{tab:table9}First two ($n_r \! = \! 0,1$) eigenvalues of 3D NPO corresponding to 
$l \! = \! 0-3$ for select $g, \lambda$ along with literature data. Taken from \cite{roy08a}.} 
\begin{ruledtabular}
\begin{center}
\begin{tabular}{ccccrl}
    &   &     &           & \multicolumn{2}{c}{Energy} \\
\cline{5-6} 
$n_r$ &  $\ell$ & $g$ & $\lambda$ & GPS \cite{roy08a}          & Literature \\ \hline
0    &  0   &  0.1    &  0.1  & 3.120081864016  & 3.1200\footnotemark[1]   \\         
1    &      &         &       & 7.231009980656  & 7.2312\footnotemark[1]   \\         
0    &  2   &         &       & 7.2439618404219 & 7.2439618404138$<$E$<$7.2439618404260\footnotemark[2],
          7.2439618404189\footnotemark[3],7.243927\footnotemark[4],7.244\footnotemark[1]  \\
1    &      &         &       & 11.317997742355 & 11.258\footnotemark[1]          \\
0    &  1   &  100    &  100  & 5.993438790399  & $-<$E$<$6.389\footnotemark[2],
5.993438873366\footnotemark[3],5.993439\footnotemark[5],5.993565\footnotemark[4],5.9936\footnotemark[1] \\
1    &      &         &       & 9.993516159965  & 9.993516\footnotemark[5],9.994694\footnotemark[4],
         9.9946\footnotemark[1]                                                              \\
0    &  3   &         &       & 9.997153638476  & 9.9969$<$E$<$10.0113\footnotemark[2],
9.997153638602\footnotemark[3],9.997145\footnotemark[5],9.9972\footnotemark[1]                           \\
1    &      &         &       & 13.99715862578  &                                      \\
\end{tabular}
\begin{tabbing}
$^{\mathrm{a}}\!$ Ref.~\cite{varshni87}. \hspace{0.3in} \= $^{\mathrm{b}} \!$ Ref.~\cite{handy93}. 
\hspace{0.3in} \= $^{\mathrm{c}} \!$ Ref.~\cite{saad06}. \hspace{0.3in}  \= 
$^{\mathrm{d}} \!$ Ref.~\cite{roy88a}. \hspace{0.3in} \= $^{\mathrm{e}} \!$ Ref.~\cite{scherrer88}. 
\end{tabbing} 
\end{center}
\vspace{-0.2in}
\end{ruledtabular}
\end{table}
\endgroup

In parallel to the works in 1D, great deal of attention has been paid to investigate 3D NPO eigenvalues, 
eigenfunctions in the past several decades, although the amount of work is visibly and surprisingly much less 
compared to the 1D counterpart. Through a super-symmetry-inspired factorization method \cite{adhikari91}, it was
possible to obtain exact algebraic-type solutions under suitable constraints on potential parameters. Right choice
of potential parameter leads to compact analytic expressions for exact eigenenergies, eigenfunctions as 
well the constraint relations. Exact solutions of the NPO in 2D, 3D have been studied in \cite{bose89} also. 
Quasi-polynomial solutions in N dimension was suggested by \cite{pons91} through the Heun confluent equation.
Some of the successful works in 3D are quoted here. Through a shifted $1/N$ expansion \cite{varshni87,roy88a}, 
results of 3D NPO were obtained for 9 sets of $n_r,l$ values for $n \! = \! 0-4$. An eigenvalue moment method \cite{handy93} 
has been quite promising in providing accurate estimates of energy bounds for the 3D NPO. A combined hypervirial 
and Pad\'e approximation \cite{witwit91} has provided very accurate results on this potential. A unified variational
treatment \cite{saad06} based on the Gol'dman and Krivchenkov Hamiltonian offered very accurate bounds. However, quite 
unfortunately, while many accurate results for 1D NPO (for example, eigenvalues accurate up to 10 decimal place were 
obtained in a number of works such as \cite{witwit96,fack86,hodgson88,witwit92a,agrawal93,stubbins95,chen98})
are available, similar results for \emph{general} states of a 3D NPO for \emph{arbitrary} sets of potential 
parameters have been obtained in very few of the works mentioned above. To the best our knowledge, such accurate results could be 
obtained by only three methods \cite{handy93,witwit91,saad06}. Even then, leaving aside \cite{witwit91}, the other two
works deal chiefly with bounds and not provide \emph{direct} eigenvalues. Thus there is a genuine lack of good-quality
results for 3D NPO. Moreover, with the rare exception of \cite{varshni87}, virtually all these above mentioned 
works have focused mainly on positive $\lambda$, even though it was known for long time that equally well-behaved 
solutions could be obtained from negative $\lambda$ provided $g \!> \!0$. Negative $\lambda$ case has been critically
examined in detail for 1D in \cite{agrawal93,stubbins95}. In the following paragraph, we will discuss the performance 
of GPS method. 

\begin{figure}
\begin{minipage}[c]{0.40\textwidth}
\centering
\includegraphics[scale=0.38]{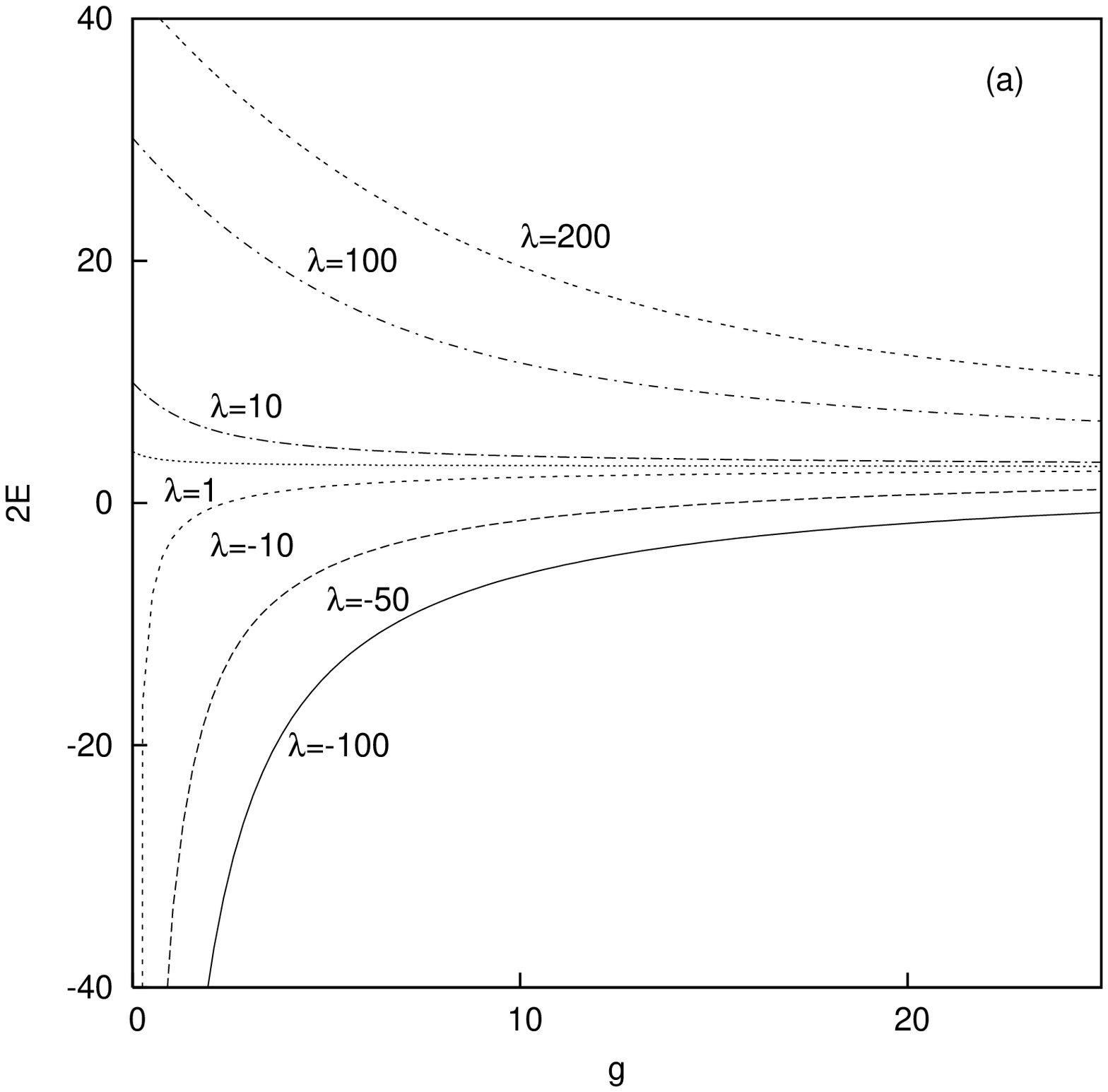}
\end{minipage}%
\hspace{0.3in}
\begin{minipage}[c]{0.40\textwidth}
\centering
\includegraphics[scale=0.38]{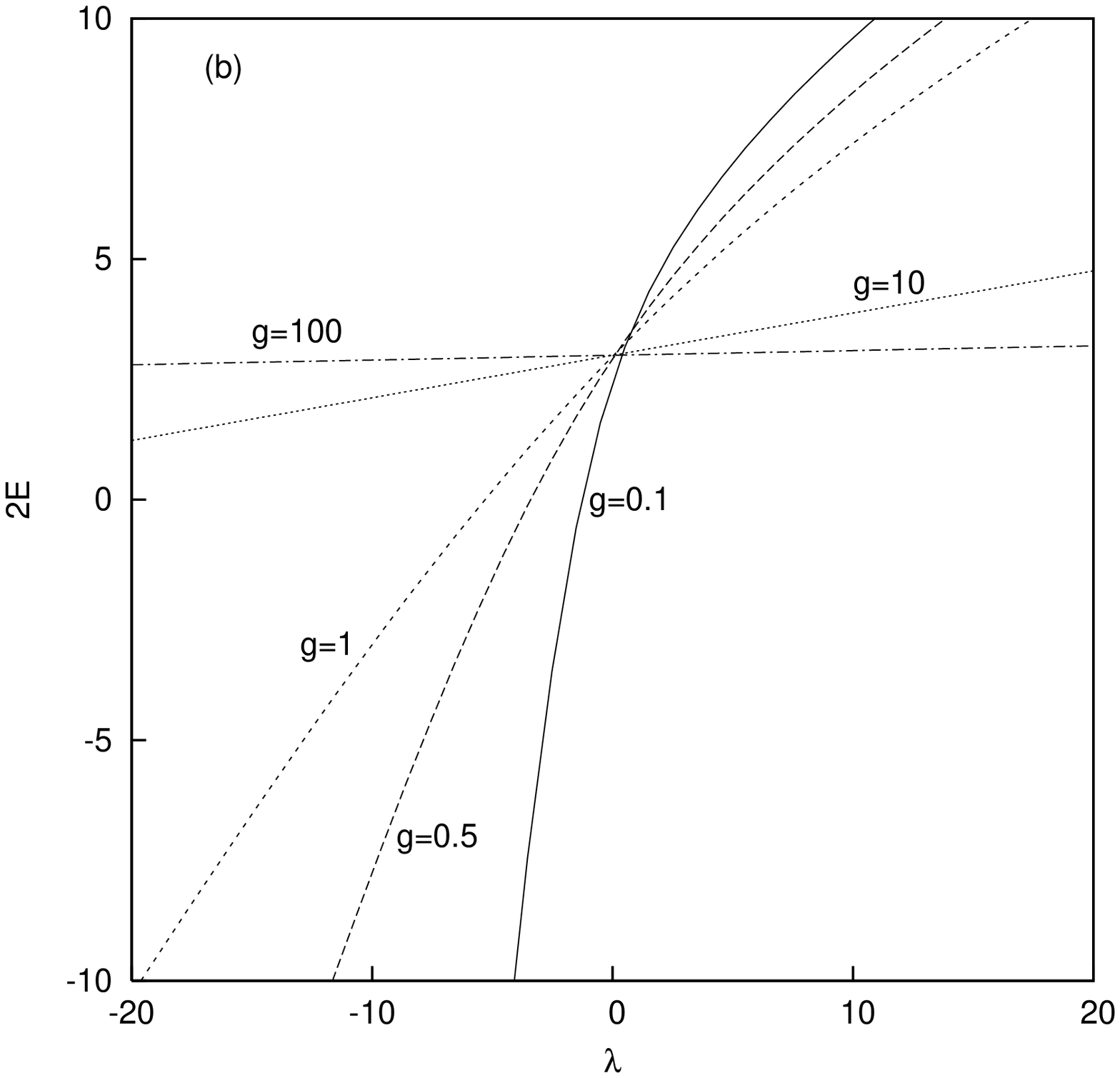}
\end{minipage}%
\caption{Ground-state energy versus (a) $g$ for fixed 
$\lambda$ (b) $\lambda$ for fixed $g$, for a 3D NPO.}
\end{figure}

Table VIII gives a few eigenvalues corresponding to certain levels of 3D NPO for some particular values of $g,\lambda$, 
which offer exact analytical solution (denoted by asterisk). These are available for $\lambda \! <0$ and presented for
lowest ($n_r\! =\! 0$) states of $l=0-2$. Literature results are quoted wherever possible. Some of them were reported long
times ago using a shifted 1/N expansion \cite{roy88a}. In all cases, 12-13 place decimal accuracy is easily achieved by 
the GPS method \cite{roy08a}. These results clearly outperform the best results available so far. Also calculations with 
$\lambda \! = \!0$ for low as well as high values lead to expected harmonic oscillator solutions promptly (up to 13th 
place of decimal). This is not presented here and can be found in \cite{roy08a}.

Next, Table IX compares the lowest two eigenstates ($n_r \! = \! 0,1$) of $l \! = \! 0-3$ with the best existing literature
data for a sufficiently broad range of interaction ($g,\lambda$ varied from 0.1--100). For small $g,\lambda$, the 
estimate of bounds obtained in eigenvalue moment method \cite{handy93} are usually good, but they deteriorate quite
badly for larger values of parameters. Significantly improved bounds have been published lately \cite{saad06}. 
It is quite clear that our method provides the most accurate direct estimates for all these states. Figure 4 vividly
displays the variation of ground-state energy against $g$ for fixed values of $\lambda$ in (a) and $\lambda$ for 
fixed values of $g$ in (b). Similar plots are obtained for higher states as well. For fixed $\lambda$, energy 
steadily decreases as $g$ increases; eventually approaching those of the harmonic oscillator asymptotically, for 
large $g$, as expected. With larger $\lambda$, this behavior is observed with a greater magnitude. Negative $\lambda$
shows a correspondingly opposite trend. In (b), changes against $\lambda$ are seen to be more prominent for smaller 
$g$; once again eigenvalues approaching harmonic oscillator energies with an increase in $g$. Many other features 
regarding higher
states (as high as up to $l \! = \! 20$) have been discussed, at length, in \cite{roy08a}. These high-lying states are 
inherently diffuse and consequently extend over a larger spatial region; hence a larger $r_{max}$ value is needed in 
order to incorporate these long-range contributions. To our knowledge, only two sets of results are available in the 
literature; one in the form of lower and upper bounds \cite{handy93}, other is from finite-difference calculations
\cite{witwit91}. Our results \cite{roy08a} practically coincide with those from latter in all cases. 

In \cite{roy08a}, a very detailed analysis was presented for energy variations with respect to interaction parameters $(g,\lambda)$. 
For this, 45 sets of $(g,\lambda)$ were chosen; $g$ values sampled as 0.1, 1, 10, 100, 1000 while $\lambda$ as $-100$, $-10$,
$-1$, $-0.1$, 0.1, 1, 10, 100, 1000. All 9 states ($1s-1g$) of $n\! = \! 0-4$  corresponding to 9 $(n_r,l)$ pairs of (+)ve 
$\lambda$ were studied in \cite{varshni87,scherrer88}, while ($-$)ve $\lambda$ has been studied rarely \cite{varshni87}. 
Our study extended to 30 states belonging to $n \! = \! 0-9$. It may be noted that in this context, the spectroscopic notation is 
more appropriate, i.e., levels are labeled as $n_r+1$ and $l$ values. So $n_r \! = \! 2, l \! = \! 1$ signifies a $3p$ 
level and so on. Interestingly it was found that for $\lambda \! > \! 0$, the ordering given as,
$$1s < 1p <2s < 1d < 2p < 1f < 3s < 2d < 1g < 3p < 2f < 1h < \cdots $$ 
was obeyed by 23 sets out of 25 sets, excepting (1,100), (10,1000). First occurrence of violation of above ordering 
was observed between the levels ($1f,3s$). Thereafter this ordering is violated in many occasions. It was also numerically 
verified that as $g$ increases in (1000,0.1) set, NPO eigenenergies approach quantum harmonic oscillator values gradually, and in the 
limit of $\lambda \! \rightarrow \! 0$, all levels belonging to a particular $n$ tend to be degenerate. Similar observations
hold true for (1000,$-$0.1) except that, now the eigenvalues approach harmonic oscillator values from below. Out of 20 
($g,\lambda$) pairs, with ($-$)ve $\lambda$, the ordering given below,  
$$1s < 1p <1d < 2s < 1f < 2p < 1g < 2d < 3s < 1h < 2f < 3p < \cdots $$ 
was satisfied by 17 sets, while (0.1,$-$10), (0.1,$-$100), (1,$-$100) sets violated this ordering. First instance of such 
violation occurs for ($2s, 1f$) and thereafter it happens for several adjacent pairs of states. Finally, it was also noted 
that, the ``usual" or most commonly observed ordering in the cases of $\lambda \! < \! 0$ and $\lambda \! > \! 0$ for 
a fixed $n$ follow a mirror-image relationship. For instance, if $n \! = \! 9$, the ordering for $\lambda \! > \! 0$ is 
$5p<4f<3h<2j<1l$ whereas for $\lambda \! < \! 0$ this is reversed, i.e., $1l<2j<3h<4f<5p$.

At this stage, a few words regarding degeneracy issues in 3D NPO are in order. It is well-known that all the ($n_r,l$)
states belonging to a particular $n$ in a 3D harmonic oscillator satisfying $n \! = \! 2n_r+l$ are degenerate. For example, there 
are 3 degenerate states corresponding to ($n_r,l$) pairs (2,0), (1,2) and (0,4), for $n \! = \! 4$. For 
$\lambda \! \ne \! 0$, such degeneracies in a 3D NPO vanishes and these are conveniently analyzed through their respective
level spacings, $\Delta E \! = \! E_{n_r,l} \! - \! E_{{n_r}',l'}$. First 4 ($n \! \leq \! 4$) such splittings of positive $\lambda$
were investigated in \cite{varshni87,scherrer88} and more recently in \cite{roy08a}, where very similar qualitative features 
were observed. However, such studies for $\lambda \! \ne \! 0$ have been made only recently \cite{roy08a} through GPS 
method; variation of 12 such splittings possible between certain adjacent levels of $n \! = \! 2-7$ were considered with
respect to the interaction parameters in potential. The first 2 splittings $E_{1,0} \! - \! E_{0,2}$, $E_{1,1} \! - 
\! E_{0,3}$ are related
to $n \! = \! 2,3$, whereas the last 3 of them $E_{1,5} \! - \! E_{0,7}$, $E_{2,3} \! - \! E_{1,5}$ and $E_{3,1} \! - \! 
E_{2,3}$ correspond to $n\! =\! 7$.
Changes in these splittings with respect to $\lambda$ were followed by varying latter from $-$0.1 to $-$100 keeping $g$ fixed at
0.1. The same for $g$ were monitored by changing latter from 1 to 1000 for fixed $\lambda \! = \! -100$. All $\Delta E$s 
tend to increase as $|\lambda|$ increases (eventually approaching a constant value in the limit of 
$\lambda \rightarrow \pm \infty$) and decrease as $g$ increases. Furthermore, the splittings tend to vanish in the limit of
large $g$ for $|\lambda|$. For more detailed discussion on energies, as well as other quantities, see \cite{roy08a}.
 
\subsection{Application to Atomic Rydberg and Hollow Resonances}
In this subsection, we briefly mention one recent application of GPS method for atomic excited states, with special
reference to singly, doubly excited Rydberg resonances in He and triply excited hollow resonances in three electron atoms. 
For this, the GPS method is used to solve the radial Kohn-Sham (KS) equation within a density functional framework. 
This approach has been very successfully applied to a broad 
range of important physical processes in atomic excitations such as multiply excited states, valence as well as core
excitations, high-lying Rydberg states, negative atoms, etc. \cite{roy02,roy04b,roy05b,roy07}. Dynamical situations have 
also been treated by this method quite well \cite{roy02a,roy02b}.
               
Triply excited atomic lithium containing all three electrons in 
$n \! \ge \! 2$ shells leaving the K shell completely empty, constitutes an interesting multi-excited atomic problem. This
is a prototypical case of a highly correlated, three-electron system under the influence of a nucleus and thus typifies the
well-known four-body Coulombic problem (an ideal system for examining delicate inter-electronic correlation). Since one-step 
photo-generation of such a state requires coherent excitation of all three available electrons, they pose significantly more 
difficulty to be produced from ground state by single-photon absorption or electron impact excitations. Besides, their close 
proximity to more than one thresholds as well presence of an infinite number of open channels offer considerable challenges 
to both experimentalists and theoreticians. A vast majority of these hollow states are auto-ionizing and have found 
important practical applications in high-temperature plasma diagnostic by means of high-resolution X-ray spectroscopy.

Development of third-generation, extreme-UV synchrotron radiation as well as availability of several powerful, 
sophisticated quantum mechanical methodologies, have inspired an overwhelming amount of work in the last three decades 
towards characterization of these states. Ever since the first electron-He scattering experiment \cite{kuyatt65}, and
$2l2l'2l''$ states in Li and highly charged ions in beam-foil experiments \cite{bruch75}, many subsequent attempts have 
been made which helped identify many bound states such as 2p$^3$ $^4$S$^o$ besides some auto-ionizing ones. However, 
the lowest 2s$^2$2p $^2$P$^o$ resonance in Li was observed in a photo-absorption spectroscopy \cite{kiernan94} through a
dual laser plasma technique. Thereafter, various higher resonances were found and tentatively classified in a wide
range of 140--165 eV. An enormous amount of experimental works have appeared in the literature lately for high-precision
determination of these resonance positions, widths, lifetimes, etc (see, for example, \cite{roy04b,roy05b}, for other
experimental reference on the subject).

Parallel to the experimental developments, an impressive amount of theoretical works have been reported in literature over 
the past years, with wide-varying range of complexity, capability, accuracy. However, due to the problems mentioned earlier, 
accurate and dependable characterization of these states has remained a formidable challenge, from a theoretical standpoint. 
Despite all these, several works are available in the literature. 
Some of the most successful methods are: (i) truncated diagonalization method \cite{conneely00} (ii) 1/Z expansion method 
\cite{safronova78} (iii) many-body perturbation theory \cite{simons79} (iv) state-specific theory \cite{nicolaides01} 
(v) configuration interaction \cite{vacek92} (vi) joint saddle point and complex coordinate rotation \cite{zhang98} (vii)
a space partition and stabilization procedure \cite{bachau96} (viii) several variants of R-matrix theory 
\cite{berrington98,zhou00} (ix) a hyperspherical coordinate approach \cite{morishita99,morishita03}, etc.

\begingroup
\squeezetable
\begin{table}
\caption {\label{tab:table10} Comparison of singly excited triplet state energies of He (a.u.).
Numbers in parentheses in column 3 denote absolute percentage errors with respect to literature data. Taken 
from \cite{roy02}.}
\begin{ruledtabular}
\begin{center}
\begin{tabular}{llll}
 State& $-$E(X-only) & $-E$(XC) & $-$E( Ref.~\cite{burgers95}) \\ \hline
$1s2s \ \ ^3 S$  & 2.17420 (2.17425\footnote{HF result, \cite{fischer77}}) & 2.17545 (0.0101) & 2.17523\\
$1s3s \ \ ^3 S$  & 2.06793   &  2.06890 (0.0102) & 2.06869\\
$1s4s \ \ ^3 S$  & 2.03606   &  2.03671 (0.0098) & 2.03651\\
$1s5s \ \ ^3 S$  & 2.02242   &  2.02264 (0.0010) & 2.02262\\
$1s7s \ \ ^3 S$  & 2.01107   &  2.01115 (0.0010) & 2.01113\\
$1s9s \ \ ^3 S$  & 2.00658   &  2.00660 (0.0000) & 2.00660\\
$1s11s \ \ ^3 S$ & 2.00431   &  2.00431 (0.0000) & 2.00431\\
$1s13s \ \ ^3 S$ & 2.00310   &  2.00310 (0.0000) & 2.00310\\
$1s15s \ \ ^3 S$ & 2.00231   &  2.00231 (0.0000) & 2.00231\\
$1s16s \ \ ^3 S$ & 2.00203   &  2.00203 &               \\
\end{tabular}  
\end{center}
\vspace{-0.1in}
\end{ruledtabular}
\end{table}
\endgroup

Density functional theory (DFT) \cite{parr89,joulbert98,dobson98,koch01} has emerged as one of the most powerful 
and successful tools for electronic structure calculation of atoms, molecules, solids in the past four decades. 
While for ground states its success was conspicuous, it was not so for excited states until only recently.  
In the last few years, a DFT-based formalism has been proposed for such resonances \cite{roy02,roy04b,roy05b,roy07}. 
This exploited a local non-variational work-function-based exchange potential \cite{harbola89}, found to be much 
more advantageous computationally compared to the non-local Hartree-Fock potential. Earlier, it was demonstrated to 
be quite successful for \emph{general} atomic excited states \cite{roy97,roy97a,singh98}. Some of the applications included: 
singly, doubly, triply excited states, low- and moderately high-lying states, valence and core excitations as well as 
auto-ionizing and satellite states, etc. However, these all used a Numerov-type finite-difference scheme for discretization
of the spatial coordinates and solution of the relevant radial KS equation, which is given as (in atomic units), 
\begin{equation}
\left[-\frac{1}{2} \nabla^2 + v_{es}(\rvec)
+v_{xc}(\rvec) \right] \psi_i (\rvec) = \varepsilon_i \psi_i (\rvec)
\end{equation}
where the three terms in left-hand side relate to kinetic, electrostatic and exchange-correlation (XC) contributions. 
Here $v_{es}(\rvec)$ contains the nuclear-attraction and classical internuclear Coulomb repulsion as,
\begin{equation}
\mathit{v}_{es}(\rvec)=-\frac{Z}{r} + \int \frac{\rho(\rpvec)}{|\rvec-\rpvec|} 
\mathrm{d}\rpvec
\end{equation}
where Z is the nuclear charge. The exchange potential is obtained through a physical interpretation, as the work 
required to move an electron against an electric field $\boldmath{\cal{E}}_x(\rvec)$ arising out of its own Fermi-hole 
charge distribution, $\rho_x(\rvec,\rpvec)$, and given by a line integral \cite{harbola89},
\begin{equation}
\mathit{v}_x (\rvec) = - \int_{\infty}^{r} 
\mbox{\boldmath $\cal{E}$}_x (\rvec) \cdot \mathrm{d} \mathbf{l}.
\end{equation}
where
\begin{equation} 
\mbox{\boldmath $\cal{E}$}_x(\rvec) = \int \frac 
{\rho_x (\rvec,\rpvec)(\rvec -\rpvec)} {|\rvec-\rpvec|^3} \ \ \mathrm{d}\rvec. 
\end{equation}
For well-defined potentials, work done must be path-independent (irrotational), which is satisfied for spherically 
symmetric systems such as those studied here. Exchange potential now can be calculated accurately as the Fermi hole 
is known exactly in terms of single-particle orbitals. Now, within the central-field approximation, 
$\psi_i(\rvec)= R_{nl}(r)\ Y_{lm}(\Omega)$. Finally, employing a suitable correlation functional (here we employ one of the
most widely used Lee-Yang-Parr potential, \cite{lee88}), one finally obtains a self-consistent set of orbitals, which 
gives the electron density as, 
\begin{equation}
\rho(\rvec)=\sum_{\mathit{i}} |\psi_{\mathit{i}}(\rvec)|^2. 
\end{equation}

However, due to the presence of Coulomb singularity at origin as well as the long-range nature of Coulomb potential, 
usual FD methods require a larger number of grid points for accurate calculation even for ground states. Hence it would be
even more difficult to capture the long tails for excited states, especially the higher ones such as Rydberg series, we 
are concerned here. In order to circumvent this problem, GPS method was invoked for accurate efficient solution of the 
KS equation in this context. In what follows we briefly mention the results on some selected states such as the 
doubly excited Rydberg states in He and hollow resonances in atomic Li, while \cite{roy02,roy04b,roy05b} could be 
referred for more complete discussion.

\begin{figure}
\centering
\includegraphics[scale=0.50,angle=270]{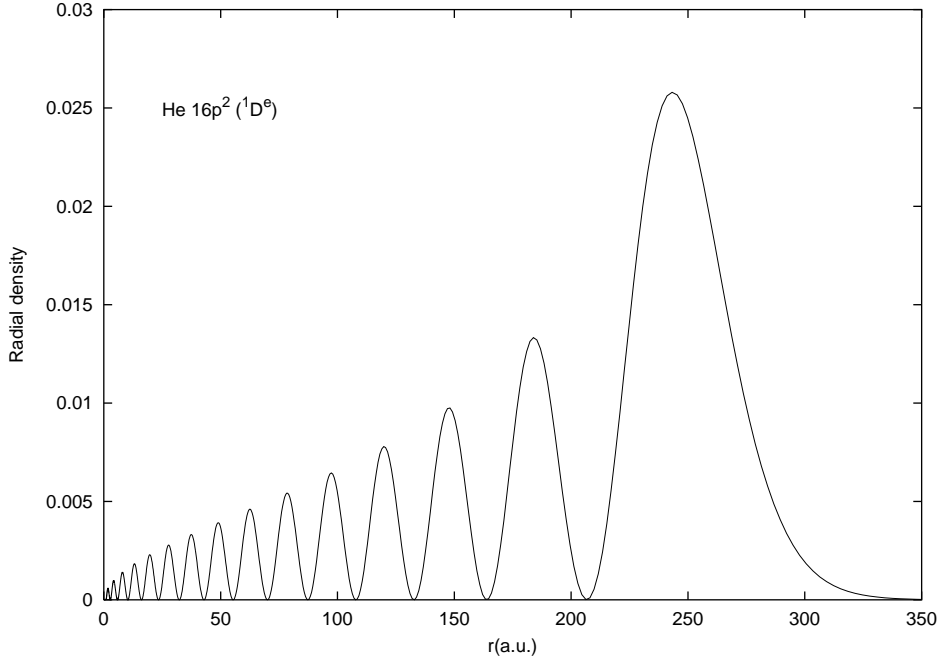}
\caption{Radial density of He 16p$^2$ $^1$D$^e$ state in a.u. Adopted from \cite{roy02}.} 
\end{figure}

Table X gives a comparison of our calculated GPS results \cite{roy02} for 1sns ($n \! = \! 2-16$) $^3$S states of He with the 
best available results in literature. Columns 2 and 3 report energies obtained from without and with correlation included.
For $n \! = \! 2$, the X-only result is only 0.0023\% above the HF value \cite{fischer77}, indicating excellent accuracy in the 
present work. For all other states (up to $n \! = \! 15$), current energies in both columns 2,3 match very closely with the much 
elaborate complex rotation calculation \cite{burgers95}. Interestingly, for high-lying Rydberg states ($n \! >\! 10$), 
energies in 3 columns (X only, XC, and reference) are seen to be essentially identical, consistent with the fact that 
for Rydberg states, asymptotic long-range Coulomb potential (that arises \emph{solely} because of the exchange potential)
remains the dominant factor for their electronic structure determination; electron correlation plays very little role.
With decrease in $n$, discrepancies in these 3 columns become noticeable as correlation now plays an increasingly 
important role. For low $n$, XC state energies have fallen slightly below the literature values, most probably due to an
overestimation caused by the correlation functional used. As $n$ goes to higher values, energy spacings decrease, and 
present work reproduces this phenomenon very well. It is worth mentioning that for high-lying states, many commonly 
used quantum mechanical methods encounter a cumbersome problem of self-consistent convergence, chiefly due to the inaccuracies
in potential and densities. However, through GPS we succeeded in getting converged results for all states as reported.
Such results were produced for even higher states by this approach in a straightforward manner (not presented here).
Finally even though the method is non-variational, anomaly in energy orderings has not been observed. Figure 5 displays the
calculated radial density for 16p$^2$ $^1$D$^e$ state of He. As expected, there are 15 maxima (first peak can be 
seen after magnification). 

\begingroup
\squeezetable
\begin{table}
\caption {\label{tab:table11}Comparison of state (in a.u.) and excitation energies (in eV) of 2s$^2$nd and 
2p$^2$ns $^2$D$^e$ resonances of Li. 1 a.u.=27.2076544 eV. See \cite{roy04b} for details.}
\begin{ruledtabular}
\begin{center}
\begin{tabular}{lllllllll}
 n  & \multicolumn{4}{c}{$\langle$A,nd$\rangle$ $^2$D$^e$}& 
      \multicolumn{4}{c}{$\langle$D,ns$\rangle$ $^2$D$^e$} \\ 
   & \multicolumn{2}{c}{$-$E(a.u.)} & \multicolumn{2}{c}{Excitation energy(eV)} 
   & \multicolumn{2}{c}{$-$E(a.u.)} & \multicolumn{2}{c}{Excitation energy(eV)} \\ 
\cline{2-3} \cline{4-5} \cline{6-7} \cline{8-9}
   & GPS & Ref. & GPS & Ref.  
   & GPS & Ref. & GPS & Ref.  \\
\hline
3  & 1.9461   & 1.9614\footnotemark[2]
   & 150.512  & 149.826\footnotemark[1],150.095\footnotemark[2]
   & 1.8458   & 1.8069\footnotemark[2] & 153.241  & 152.364\footnotemark[1],154.299\footnotemark[2]    \\
7  & 1.8995   & 1.9084\footnotemark[2] &  151.780 
   & 151.069\footnotemark[1],151.537\footnotemark[2]
   & 1.7618   & 1.7607\footnotemark[2] &  155.526 
   & 154.895\footnotemark[1],155.556\footnotemark[2] \\
9  & 1.8951   & 1.9042\footnotemark[2] &  151.899 
   & 151.185\footnotemark[1],151.652\footnotemark[2] 
   & 1.7564   & 1.7571\footnotemark[2] &  155.673 
   & 155.026\footnotemark[1],155.654\footnotemark[2] \\
10 & 1.8936   & 1.9029\footnotemark[2] &  151.940
   & 151.217\footnotemark[1],151.687\footnotemark[2] 
   & 1.7549   &       & 155.714    & 155.065\footnotemark[1]     \\
11 & 1.8924   & 1.9020\footnotemark[2] &  151.973 
   & 151.241\footnotemark[1],151.712\footnotemark[2]
   & 1.7538   &       & 155.744    & 155.102\footnotemark[1]     \\
12 & 1.8914   &       & 152.000    & 151.260\footnotemark[1]
   & 1.7530   &       & 155.765    & 155.122\footnotemark[1]     \\
16 & 1.8893   &       & 152.057    & 151.302\footnotemark[1]
   & 1.7512   &       & 155.814    & 155.169\footnotemark[1]     \\
20 & 1.8884   &       & 152.082    & 151.321\footnotemark[1]
   & 1.7504   &       & 155.836    & 155.191\footnotemark[1]     \\
22 & 1.8882   &       & 152.087    & 151.327\footnotemark[1]
   & 1.7501   &       & 155.844    & 155.197\footnotemark[1]     \\
24 & 1.8880   &       & 152.094    &
   & 1.7499   &       & 155.850    & 155.202\footnotemark[1]     \\
25 &          &       &            &
   & 1.7498   &       & 155.853    & 155.205\footnotemark[1]     \\
\end{tabular}
\begin{tabbing}
$^{\mathrm{a}}\!$ Ref.~\cite{zhou00}. \hspace{0.3in} \= $^{\mathrm{b}} \!$ Ref.~\cite{conneely02}. 
\end{tabbing} 
\end{center}
\vspace{-0.3in}
\end{ruledtabular}
\end{table}
\endgroup

Next, some selected hollow states of Li, obtained from GPS method, are reported in Table XI, along with some literature data, 
in terms of non-relativistic term energies as well as excitation energies. In the literature, usually, only the latter is
reported. In a DFT study on low-lying singly excited states of open-shell atoms \cite{singh98}, it was observed that 
excitation energies from X-only and numerical HF calculations show good agreement with each other. However, excitation
energies did not show significant improvements upon inclusion of two correlation energy functionals (one relatively 
simple local Wigner and the other considerably involved non-local LYP), although excited-state energies were improved
significantly. In this table, they both are presented. The latter is estimated with respect to the accurate ground-state
energy of Li, $-7.47805953$ a.u. \cite{chung91}, calculated from a full core plus correlation via a multi-configuration
interaction wave function, in order to maintain consistency with literature. The present DFT calculation yields the same
as $-7.4782839$ a.u. For convenience, an independent particle model classification \cite{conneely00,conneely02} is 
adopted; thus the six core Li$^+$ n=2 intra-shell doubly excited states, \emph{viz.}, 2s$^2$ $^1$S$^e$, 2s2p $^3$P$^o$,
2p$^2$ $^3$P$^e$, 2p$^2$ $^1$D$^e$, 2s2p $^1$P$^o$ and 2p$^2$ $^1$S$^e$ are denoted by A,B,C,D,E and F respectively. This
table compares the even-parity $\langle \mathrm{A,nd} \rangle$ and $\langle \mathrm{D,ns} \rangle$ $^2$D$^e$ Li hollow 
states with literature data, arising from electronic configurations 2s$^2$nd and 2p$^2$ns having n up to 24,25 
respectively. For the first series, to the best of our knowledge, no experimental results are available, whereas for 
latter, only the lowest state is detected experimentally at 144.77 eV in photo-electron spectroscopy \cite{cubaynes96}. DFT
excitation energy matches excellently with the experimental value (only 0.043 eV lower with a deviation of 0.03\%). The
term energies are slightly underestimated in all cases with respect to the truncated diagonalization result 
\cite{conneely02}, whereas for latter, some overestimation is noticed for n=2,4--7, which could occur either due to (a)
non-variational nature of the exchange potential employed and/or (b) inadequacy of LYP correlation energy functional. 
Higher resonances (up to n=22,25) for the two series have been theoretically investigated through the R-matrix method 
\cite{zhou00}; DFT excitation energies show discrepancies in the range of 0.46--0.50\% and 0.04--0.58\% for them. Finally Fig.~6 
depicts radial densities for some selective hollow states of Li, which show the characteristic shell structures through
superposition of radial densities. 

\begin{figure}
\begin{minipage}[c]{0.40\textwidth}
\centering
\includegraphics[scale=0.45]{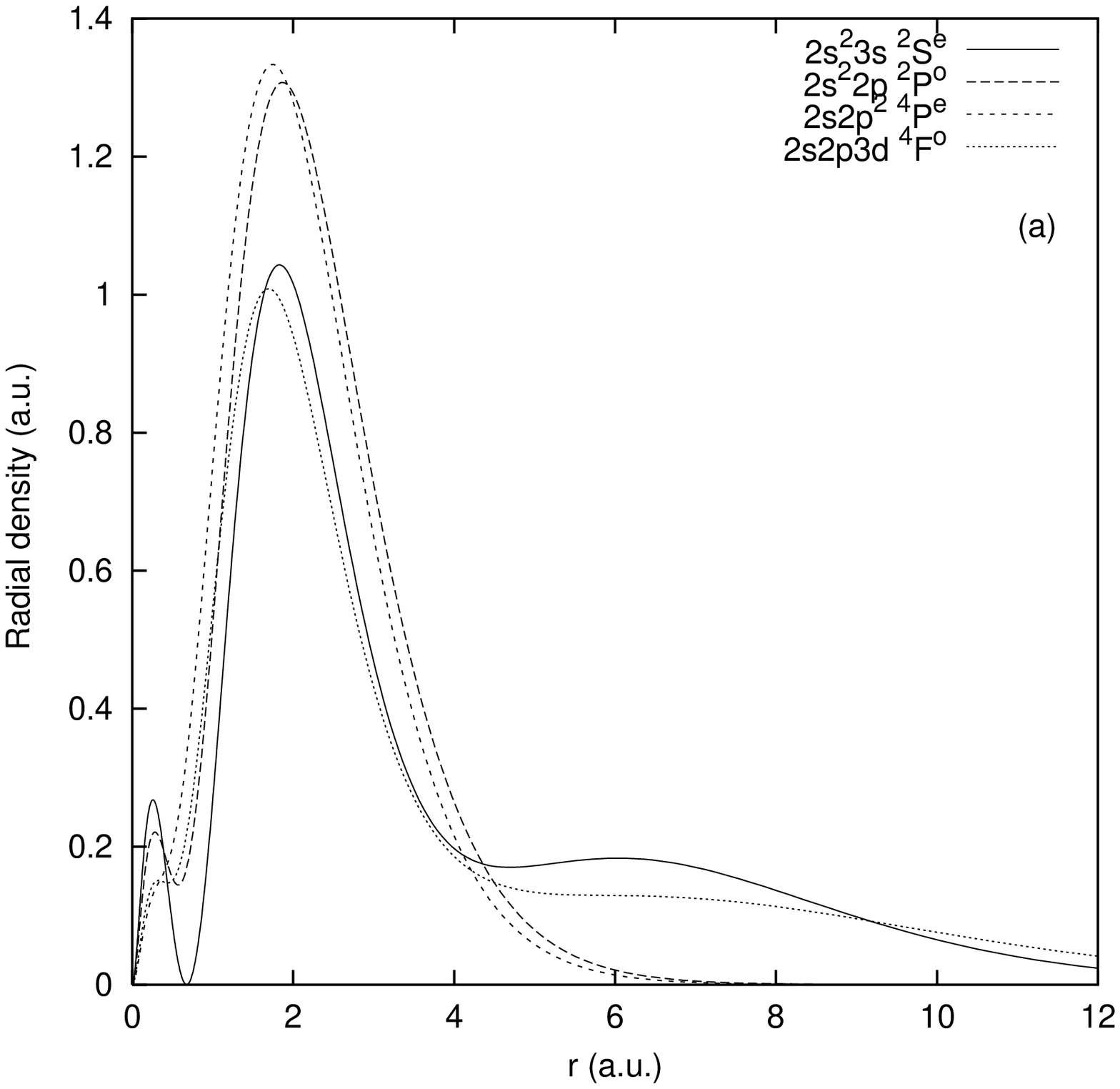}
\end{minipage}%
\hspace{1in}
\begin{minipage}[c]{0.40\textwidth}
\centering
\includegraphics[scale=0.45]{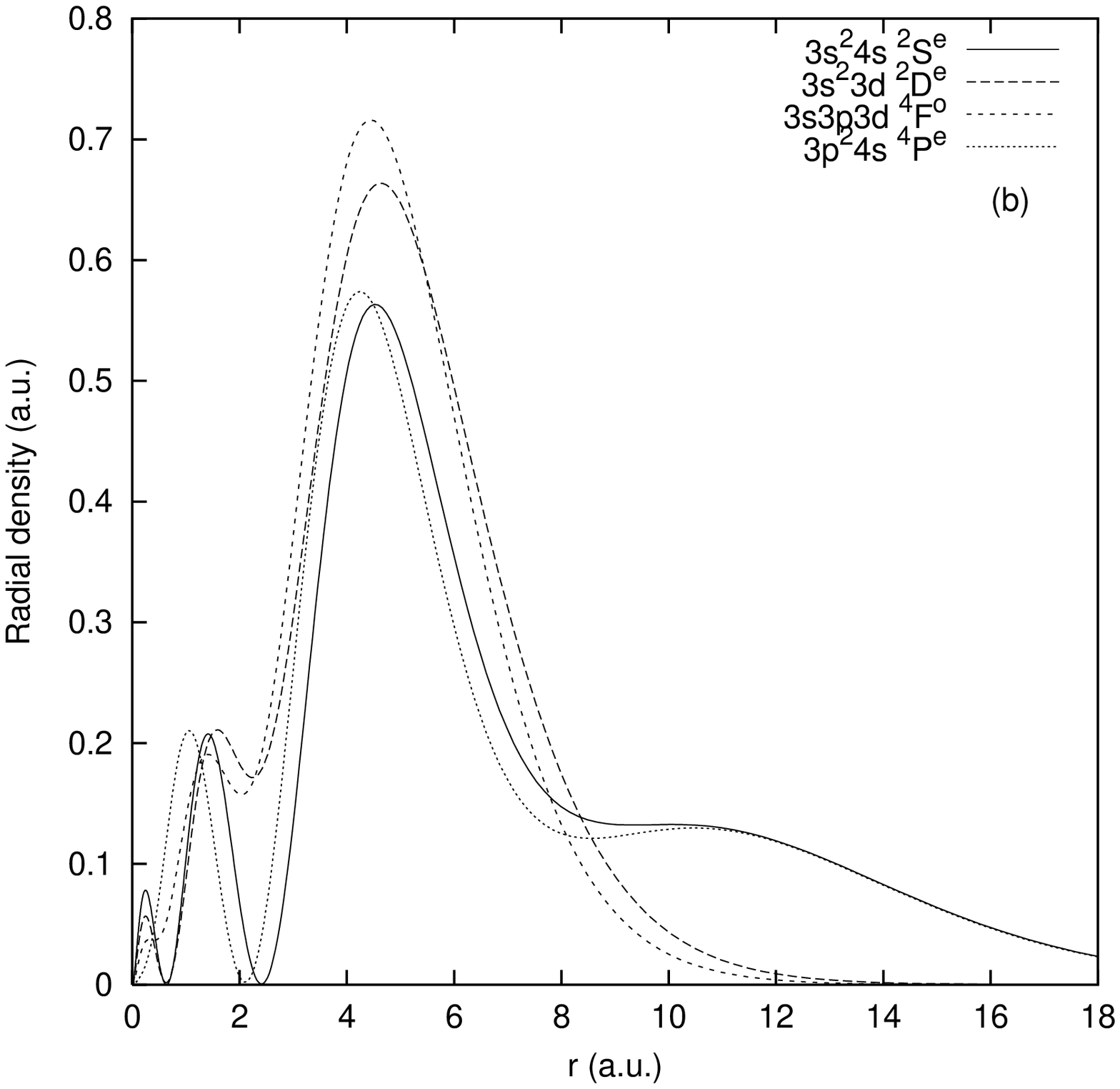}
\end{minipage}%
\caption{The radial densities (a.u.) of Li for (a) 2s$^2$3s $^2$S$^e$, 2s$^2$2p $^2$P$^o$, 2s2p$^2$ $^4$P$^e$, 
2s2p3d $^4$F$^o$ and (b) 3s$^2$4s $^2$S$^e$, 3s$^2$3d $^2$D$^e$, 3s3p3d $^4$F$^o$, 3p$^2$4s $^4$P$^e$ states respectively.
Taken from \cite{roy04b}.}
\end{figure}

In \cite{roy04b}, DFT calculations have also been reported for twelve $2l2l'$n$l''$ (n$\geq$2) triply excited hollow resonance 
series of Li, {\em viz.,} 2s$^2$ns $^2$S$^e$, 2s$^2$np $^2$P$^o$, 2s$^2$nd $^2$D$^e$, 2p$^2$ns $^2$D$^e$,$^4$P$^e$, 
2s2pns $^4$P$^o$, 2s2pnp $^4$D$^e$, 2p$^2$np $^2$F$^o$,$^4$D$^o$, 2p$^2$nd $^2$G$^e$, $^4$F$^e$ and 2s2pnd $^4$F$^o$, 
covering a total of about 270 low-, moderately high- and high-lying states. Except for one state, the discrepancy
in state energies remained well within 0.98\%, whereas GPS excitation energies deviated by 0.02--0.58\% with respect to 
the literature results. As a further extension, additional calculations were also performed for 37 $3l3l'$n$l''$ 
(n$\geq$3) doubly hollow resonances {\em viz.,} $3l3l'$n$l''$(3$\le$n$\le$6) ($^2$S$^e$, $^2$P$^o$, $^2$D$^e$, $^2$F$^o$, 
$^4$S$^o$, $^4$P$^{e,o}$, $^4$D$^{e,o}$ and $^4$F$^o$) (seven are n=3 intra-shell type) of Li with both K, L shells empty 
(up to n=6) in the photon energy range 175.63--180.51 eV, having different symmetries and multiplicities. In this occasion 
also, DFT calculation shows quite good agreement with recent literature data. In this case, however, literature data is far more
sparse compared to the hollow resonances. The most distinctive feature of these are: (a) they are weaker by about an order
of magnitude, compared to their hollow cousins (b) broad and having much larger widths \cite{azuma97}. The major difficulty
in handling such hollow resonances at higher photon energies arises mainly due to a very rapid increase in the density 
of triply and other lower excited states of same symmetry, as well as of large number of available open channels, leading
to very strong, complicated correlation effects. Nevertheless, some attempts have been made to study these states. Some of the
prominent theoretical works include: (a) complex scaling method having correlated basis functions constructed from B splines 
\cite{madsen00} (b) state specific theory \cite{piangos03}, etc. For more detailed discussion as well as available
experimental and theoretical works on these, see \cite{roy04b}. 
Hollow resonances of Li-isoelectronic series have also been studied successfully by this method in \cite{roy05b}, where 
8 $2l2l'$n$l''$ (3 $\leq$n$\leq$6) hollow resonance series, namely, 2s$^2$ns $^2$S$^e$, 2s$^2$np $^2$P$^o$, 2s$^2$nd 
$^2$D$^e$, 2p$^2$ns $^2$D$^e$, 2s2pns $^4$P$^o$, 2s2pnp $^4$D$^e$, 2p$^2$np $^4$D$^o$, of all the 7 positive ions from 
Z=4--10 were reported.

\section{Concluding Remarks}
Quantum mechanics has nowadays spread applications widely in almost every imaginable area in contemporary science and 
technology, including physics, chemistry, materials science, nanoscience, etc. The focus of this chapter, central potentials,
especially those \emph{singular} or near-singular, take the centrestage in atomic, molecular, optical physics and chemistry.
However, any exactly solvable quantum system, more so for the singular potentials, are very scarce. Innumerable attempts 
have been made to solve the respective Schr\"odinger equation almost ever since the early inception of quantum mechanics. 
While variational and perturbative approaches remain the most commonly employed, many other attractive, elegant formalisms 
are available these days, which offer physically meaningful and quite accurate results for various quantum mechanical systems of 
interest. An enormous number of analytical, semi-numerical, purely numerical techniques have also been developed over the 
decades for this purpose. Nevertheless, as discussed above, there is still a great need for better approaches. Because, 
for all the physical systems covered in this work (and probably also true for many other situations not touched upon 
here) many of these methods would be satisfactory for certain ranges of potential parameters and less successful for
other sets. Moreover, a vast majority of these methods work well for lower states; outstanding difficulties and 
challenges are encountered for higher states. Additionally extraction of radial density, as well as other expectation
values are not straightforward task. Very few methods would satisfy \emph{all} these criteria. In essence, there is a great need for a methodology which can satisfy all these
criteria. 

Here we have presented an account of the development of GPS method in the context of central (both singular and non-singular) 
potentials in quantum 
mechanics. Motivation, background, need for such a method as well its details have been discussed at some length. Although
initially designed for Coulombic singular systems, its success for other singular systems as well as for other
non-singular central potentials was realized promptly. The formalism is quite \emph{general}, in the sense that it delivers
uniformly accurate results for \emph{lower and higher} states of a \emph{broad} spectrum of potentials (describing a variety
of physical systems) covering a \emph{wide} range of interaction/coupling. Its usefulness and applicability was 
demonstrated for some specific potentials in Section IV, such as spiked oscillator, Hulth\'en, Yukawa, power-law, logarithmic,
non-polynomial oscillator, and lastly, some Rydberg and hollow resonances in atomic systems, etc. It has been successfully 
applied to certain other systems as well,
which have not been mentioned at all (such as Hellmann potential or Coulomb potential perturbed by a linear and
quadratic coupling) and these could be found in 
refs~\cite {roy02,roy02a,roy02b,roy04,roy04a,roy04b,roy05,roy05a,roy05b,sen06,roy07,roy08,roy08a,roy08b}. Eigenvalues,
eigenfunctions, radial densities, spatial expectation values are obtained in a simple manner. A comparison with literature
data reveals that in most cases, results obtained are quite competitive to the best ones or surpasses the accuracy of 
existing best results available. In almost all cases, it helps to estimate many new states for the first time, which could 
constitute useful references for future investigations. Finally, it is hoped that this method will continue to remain as a 
valuable tool for many other physical systems in the future.

\section{acknowledgments}
I express my sincere gratitude and thanks to Editor, Prof.~F.~Columbus, for giving this opportunity to present my 
work in this exciting area in this edition. I am very much thankful to Nova Publishers, NY, USA, for their generous support
and extending deadline for submission. Valuable discussion with Prof.~P.~Panigrahi is gratefully acknowledged.
It is a pleasure to thank my colleagues at IISER, Kolkata, for providing a friendly working environment.  

\bibliographystyle{unsrt}
\bibliography{refn}
\end{document}